\documentclass[usenatbib, useAMS, referee]{biom}
\usepackage{amsfonts,amsmath,amssymb,bbm,bm}
\usepackage{graphicx,psfrag,epsf}
\usepackage{enumerate}
\usepackage{textalpha, pifont}
\newcommand{\cmark}{\ding{51}}%
\newcommand{\xmark}{\ding{55}}%
\usepackage{soul}
\usepackage[normalem]{ulem}
\usepackage{subcaption,lscape}
\usepackage[font=small,labelfont=bf]{caption}
\usepackage{float,multirow,booktabs}
\usepackage[table,xcdraw]{xcolor}
\usepackage{url} 
\usepackage{rotating,titlesec,enumitem}
\usepackage{algorithm,tabularx}
\usepackage[noend]{algpseudocode}
\usepackage[utf8]{inputenc}
\usepackage[english]{babel}

\newcounter{algsubstate}

\makeatletter
\newcommand{\multiline}[1]{%
	\begin{tabularx}{\dimexpr\linewidth-\ALG@thistlm}[t]{@{}X@{}}
		#1
	\end{tabularx}
}
\makeatother

\usepackage{NotationCommands}

\newcommand{\transp}{{\sf T}}

\title[Robust Bayesian Graphical Regression]{Robust Bayesian Graphical Regression Models for Assessing Tumor Heterogeneity in Proteomic Networks}
\author{Tsung-Hung Yao$^{1,*}$\email{yaots@umich.edu}, 
Yang Ni$^{2}$, and Anindya Bhadra$^{3}$, and Jian Kang$^{1}$, \\
\bf{and Veerabhadran Baladandayuthapani}$^{1,**}$\email{veerab@umich.edu}\\
$^{1}$Department of Biostatistics, University of Michigan at Ann Arbor, Ann Arbor, Michigan, U.S.A. \\
$^{2}$Department of Statistics, Texas A\&M University, College Station, Texas, U.S.A. \\
$^{3}$Department of Statistics, Purdue University, West Lafayette, Indiana, U.S.A.}

\begin{document}
\date{{\it Received October} 2023. {\it Revised February} 2024.  {\it Accepted March} 2024.}
\pagerange{\pageref{firstpage}--\pageref{lastpage}} 
\volume{64}
\pubyear{2024}
\artmonth{December}
\doi{10.1111/j.1541-0420.2005.00454.x}
\label{firstpage}
	
\begin{abstract}
Graphical models are powerful tools to investigate complex dependency structures in high-throughput datasets. However, most existing graphical models make one of the two canonical assumptions: (i) a homogeneous graph with a common network for all subjects;  or (ii) an assumption of normality especially in the context of Gaussian graphical models. Both assumptions are restrictive and can fail to hold in certain applications such as proteomic networks in cancer. To this end, we propose an approach termed robust Bayesian graphical regression (rBGR) to estimate heterogeneous graphs for non-normally distributed data. rBGR is a flexible framework that accommodates non-normality through random marginal transformations and constructs covariate-dependent graphs to accommodate heterogeneity through graphical regression techniques. We formulate a new characterization of edge dependencies in such models called conditional sign independence with covariates along with an efficient posterior sampling algorithm. In simulation studies, we demonstrate that rBGR outperforms existing graphical regression models for data generated under various levels of non-normality in both edge and covariate selection. We use rBGR to assess proteomic networks across two cancers: lung and ovarian, to systematically investigate the effects of immunogenic heterogeneity within tumors. Our analyses reveal several important protein-protein interactions that are differentially impacted by the immune cell abundance; some corroborate existing biological knowledge whereas others are novel findings. \\

\end{abstract}


\maketitle
	
\section{Introduction}


Graphical models are ubiquitous and powerful tools to investigate complex dependency structures in high-throughput biomedical datasets such as genomics and proteomics \citep{Intro_PGM}. They allow for holistic exploration of biologically relevant patterns that can be used for deciphering biological processes and formulating new testable hypotheses. 
However, most existing graphical models make one of the two canonical assumptions: (i) a homogeneous graph that is common to all subjects; 
or (ii) an assumption of normality as in the context of Gaussian graphical models \citep{Ni2022}. 
However, in some biomedical applications 
such as inference of proteomic networks in cancer, both assumptions are violated, as we illustrate next. 



\begin{figure}
    \centering
    \includegraphics[width=\textwidth]{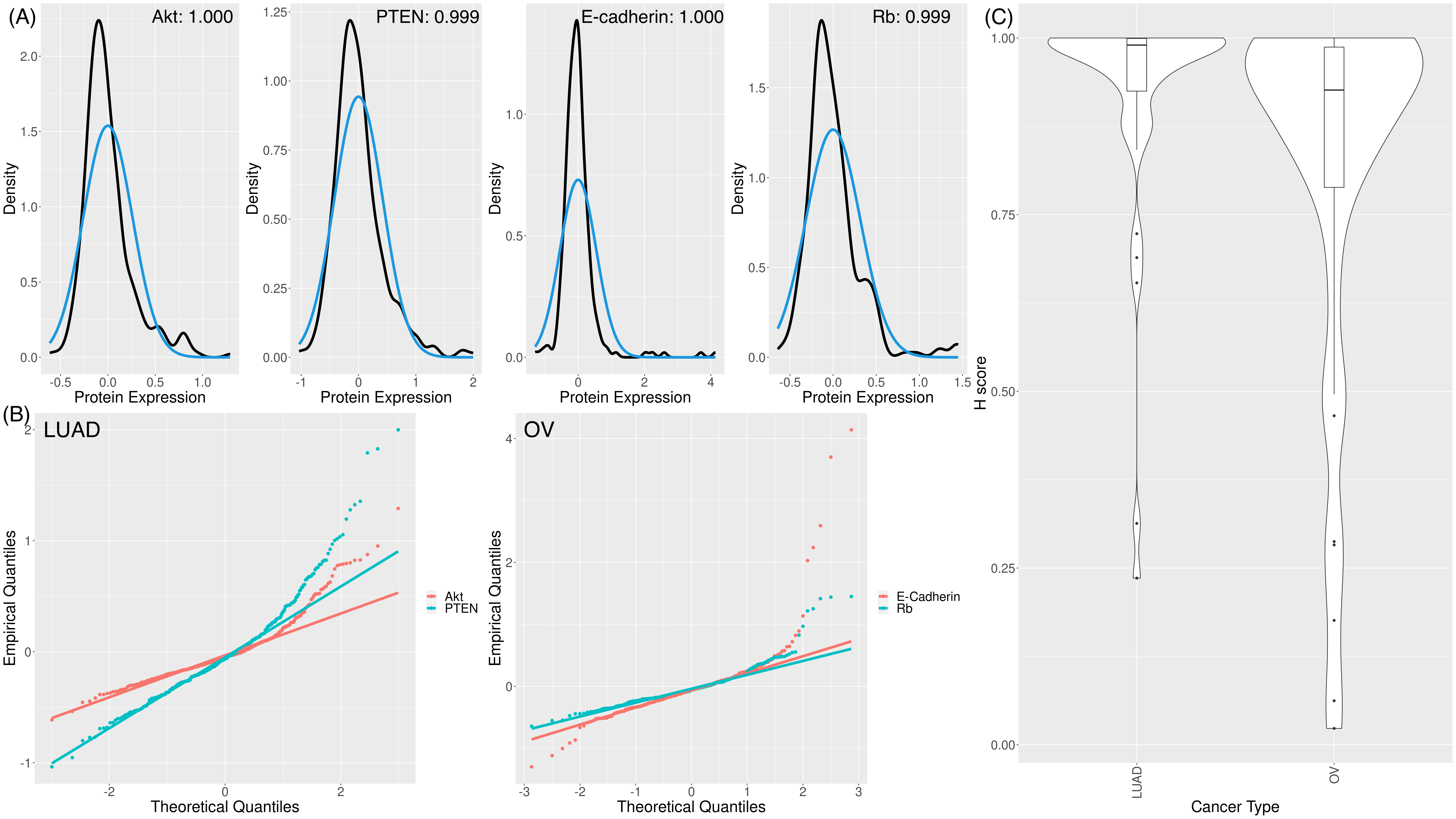}
    \caption{Non-normality levels of protein expression data from lung adenocarcinoma (LUAD) and ovarian cancer (OV). The empirical density plots from actual data (black) and the normal distribution (blue) for the expression of four proteins with the H-score are shown in the upper-right of each protein in Panel (A). 
    Panel (B) illustrates the expression of four proteins in LUAD (Akt and PTEN) and OV (E-Cadherin and Rb) with qq-plots. Panel (C) shows the violin plot of H-scores across all proteins in LUAD and OV cancers. The H-score is bounded between zero and one, and a higher H-score implies a higher level of non-normality (see manuscript for description). }
    \label{fig:RPPA_HScore}
\end{figure}

\noindent {\it Proteomic networks and tumor heterogeneity.}
Proteins control many fundamental cellular processes through a complex but organized system of interactions, termed protein-protein interactions (PPIs) \citep{pmid32228680}. Moreover, aberrant PPIs are associated with various diseases including cancer, and investigating PPI can lead to effective strategies and treatments, including immunotherapies, tailored to different individuals \citep{pmid32228680}. Consequently, it is highly desirable to elucidate PPIs in cancer and construct flexible graphical models that can identify multiple types and ranges of dependencies. 
Modern data collection methods have allowed systematic assessment of multiple proteins simultaneously on the same tumor samples, often referred to as high-throughput proteomics \citep{pmid26246866}. However, the resulting data are typically not normally distributed even after extensive preprocessing and data transformations (e.g. logarithmic). As an illustration, Figure \ref{fig:RPPA_HScore} shows the level of non-normality in protein expression data after logarithmic transformation for two cancers: lung adenocarcinoma (LUAD) and ovarian cancer (OV) samples from The Cancer Genome Atlas (TCGA) \citep{pmid24071849} that are used as case-studies in this paper. Panels (A) and (B) display the empirical density and the quantile-quantile (q-q) plots of four exemplar proteins: Akt and PTEN for LUAD, and E-Cadherin and Rb for OV. Both the empirical distributions and q-q plots demonstrate deviations from normal distribution with heavier tails as shown in Panels (A) and (B). The level of non-normality is further quantified using the H-score, defined as $H(\bY)=2\Phi[\log\{1-\text{pval}(\bY)\}]$, 
where $\Phi$ is the cumulative distribution function of the standard normal distribution, and $\text{pval}(\bY)$ is the p-value of the Kolmogorov-Smirnov test for the normality of $\bY$ \citep{RCGM}. The H-score is bounded between zero and one, and a higher H-score implies increased departure from normality. The H-scores for all four proteins are $>0.999$, consistent with the conclusions from the empirical and q-q plots. Panel (C) shows the H-score across \underline{all} the proteins in our datasets, indicating a high degree of non-normality across both cancers.

Another axes of complexity that arises in cancer research is {\it tumor heterogeneity}. It is now well-established that tumors are heterogeneous with distinct proteomic aberrations even for the same type of cancer across different patients \citep{pmid24587830}. Accumulating evidence suggests that considering tumor heterogeneity, in general, and specifically at the level of PPIs can enhance our understanding of tumorigenesis and the development of anti-cancer treatments \citep{pmid32228680}. Specifically, tumor heterogeneity differentially impacts the PPIs across different patients and results in varied treatment responses \citep{pmid32228680}. Hence, incorporating patient-specific information, i.e., accounting for tumor heterogeneity, could provide valuable clues to identify PPIs disrupted during carcinogenesis. 
In summary, constructing PPI networks poses two main statistical challenges simultaneously: (i) coherently accounting for non-normality in proteomic networks, and (ii) incorporating heterogeneous patient-specific information in graphical modeling.

\vspace{-3.5mm}
\vspace{-3.5mm}

\noindent {\it Existing methods and modeling background.}~~
Most existing methods address the aforementioned challenges separately, i.e., either accommodating non-normality without accounting for the sample-specific information \citep[e.g.][]{10.2307/20441306, 10.1214/10-AOAS397} or requiring normality when incorporating patient-specific information \citep{Ni2022}.  
To accommodate non-normality, existing approaches transform the original variables into normal variables either via deterministic functions \citep[e.g.][]{10.1214/10-AOAS397,Liu12_NonParaNorm, chung2022phylogenetically} or via random transformations \citep[e.g.][]{tlasso, robustTGraph}. For instance, \citet{CSI_Anindya} use a Gaussian scale mixture technique that generalizes the $t$-distribution and introduce a new graph characterization for undirected graphs. \citet{RCGM} further generalize this concept to characterize chain graphs with both directed and undirected edges. However, all existing models mentioned above assume a common graph across all patients and fail to incorporate the subject-specific information. 

More recently, several studies incorporate the subject-specific information under explicit Gaussian assumptions. Multiple Gaussian graphical models were first proposed to estimate graphs that vary across heterogeneous sub-populations \citep[e.g.][]{pmid19881892, pmid24817823, pmid26078481}. \citet{BGR_Normal} introduce a more general framework called ``graphical regression" that constructs covariate-dependent graphs through a regression model and incorporates both continuous and discrete covariates, in directed as well as undirected settings \citep{GGMx}. Similarly, \citet{RegGMM} provide a penalized procedure to estimate undirected graphs through Gaussian graphical regression and introduce continuous covariates in both the mean and the covariance structures.  
However, all these models are developed under the normality assumption for inferential and computational reasons. To the best of our knowledge, no existing method incorporates subject-specific information under non-Gaussian settings and motivates development of new methodology. We summarize six important and relevant models mentioned above in Table \ref{tab:litRev} and compare these models in four different aspects.

\vspace{-3.5mm}
\begin{table}
     \caption{Comparison of existing methods and our proposed rBGR method across four different properties.}
     \centering
     \setlength\extrarowheight{-7pt}
     \label{tab:litRev}
     \begin{tabular}{l | cccc}
     \multirow{2}{*}{Method} & Uncertainty & \multirow{2}{*}{Undirected} & Sample- & \multirow{2}{*}{Non-Normality} \\
      & Quantification &  & specific & \\
     \midrule
     GGMx \citep{GGMx} & \cmark & \cmark & \cmark & \xmark \\
     RegGMM \citep{RegGMM} & \xmark & \cmark & \cmark & \xmark \\
     GSM \citep{CSI_Anindya} & \cmark & \cmark & \xmark & \cmark \\
     BGR  \citep{BGR_Normal} & \cmark & \xmark & \cmark & \xmark\\
     RCGM  \citep{RCGM} & \cmark & \cmark & \xmark & \cmark \\
     rBGR (the proposed) &  \cmark & \cmark & \cmark &  \cmark \\
     \bottomrule
     \end{tabular}
\end{table}
\vspace{-3.5mm}

\noindent To address these challenges simultaneously, we develop a unified and flexible modeling strategy called \ul{robust Bayesian graphical regression} (rBGR), which allows construction of subject-specific graphical models for non-normally distributed continuous data. rBGR makes three main contributions:
\begin{itemize}[noitemsep,topsep=0pt]
    \item[(a)] {\it Robust framework to build subject-specific graphs for non-normal data.} rBGR robustifies the normal assumption via random transformation and incorporates covariates employing graphical regression strategies. By accommodating non-normality via random scale transformations, we obtain a Gaussian scale mixture, which presumes an underlying latent Gaussian variable, allows explicit incorporation of covariates in the precision matrix (Section \ref{sec:rBGR_multiNorm}), and admits efficient posterior sampling procedures (Section \ref{sec:inference}). 
    \item[(b)] {{\it New characterization of dependency structures for non-normal graphical models}. The introduction of the random marginal transformations engenders a new type of edge characterization of the conditional dependence for non-normal data, called conditional sign independence with covariates (CSIx, Proposition \ref{Prop_FuncCSI}). CSIx is a generalization of the notion of conditional sign independence (CSI), introduced by \citet{CSI_Anindya}, which explicitly characterizes the sign dependence between two variables that holds for a much broader class of models than Gaussian graphical models. 
    We demonstrate via multiple simulations that rBGR can accurately recover dependency structures under different levels of non-normality and against competing graphical regression approaches that assume normality (Section \ref{sec:simulation}).}
    \item[(c)] {\it Deciphering impact of immunogenic heterogeneity in proteomic networks.} We use rBGR to assess proteomic networks across two cancers, lung and ovarian, to systematically investigate the effects of the inherent immunogenic heterogeneity within tumors. Specifically, we quantify immune cell abundance across tumors and build PPI networks that vary across different immune cell abundance. Our analyses reveal several important hub proteins and PPIs that are impacted by the immune cell abundance; some corroborate existing biological knowledge whereas others are novel associations 
    (Section \ref{sec:realData}).
\end{itemize}

The rest of the paper is organized as follows: we introduce rBGR models and characterization in Section \ref{sec:rBGR}. Section \ref{sec:estimation} focuses on priors and estimation, and Section \ref{sec:inference} delineates the posterior inference via Gibbs sampling. 
In Section \ref{sec:simulation}, we conduct a series of simulations to evaluate the operating characteristics of rBGR against competing approaches. Section \ref{sec:realData} provides a detailed analysis of the TCGA dataset, results, biological interpretations, and implications. The paper concludes by discussing implications of the findings, limitations, and future directions in Section \ref{sec:Diss}. A general purpose \texttt{R} package and datasets used in this paper for constructing PPI networks is also provided at \url{https://github.com/bayesrx/rBGR}. 



\section{Robust Bayesian Graphical Regression (rBGR)}\label{sec:rBGR}
We start by reviewing the Gaussian graphical regression (Section \ref{sec:GaussianGraphReg}), which is a special case of rBGR under the normality assumption, and then generalize it to the robust case by random transformations (Section \ref{sec:rBGR_multiNorm}). Subsequently, the introduction of the random scale changes the interpretation of graph and motivates a new edge characterization (Section \ref{sec:char_CSIx}).

\subsection{Gaussian Graphical Regression}\label{sec:GaussianGraphReg}
Consider $p$-dimensional random vectors $\bY_i=(Y_{i1},\ldots, Y_{ip})^\transp\in\mathbb{R}^p$ as (continuous) responses with $q$-dimensional random vectors of $\bX_i=(X_{i1},\ldots, X_{iq})^\transp\in\mathbb{R}^q$ as covariates for subject $i=1,\ldots,n$. A subject-specific PPI network from proteomics data $\bY_i$ is constructed to vary based on the immune cell abundance $\bX_i$ (Section \ref{sec:realData}).
Let $G_i=(V,E_i)$ be an undirected graph over $p$ nodes, where $V=\{1,\ldots,p\}$ is the set of nodes representing $\bY_i$ and $E_i\subset V\times V$ is the set of undirected edges in the graph for subject $i$. An undirected edge exists between nodes $j$ and $k$ if $\{j,k\}\in E_i$. 
Under Gaussian assumption, given the covariates $\bX_i$, suppose $\bY_i$ follows a multivariate normal distribution,
\begin{align}\label{eq:Gau_MultiNorm}
    \bY_i \mid \bX_i \sim \bN_p({\bf 0}, \tilde{\bOmega}^{-1}(\bX_i)), \textrm{ for } i=1,\ldots,n,
\end{align}
where $\tilde{\bOmega}(\bX_i)=\{\tilde{\omega}^{j,k}(\bX_i)\}_{p\times p}, j,k\in V$ 
is a functional precision matrix (of covariates) with each element $\tilde{\omega}^{j,k}(\bX_i)$ as a function that depends on $\bX_i$. The functional precision matrix characterizes the graph $G_i$ through zero precision elements. Specifically, a zero element of the precision matrix represents a missing edge in the graph, e.g., for the case of scalar precision matrix, $\tilde{\omega}^{j,k}(\bX_i)=\tilde{\omega}^{j,k}$, zero precision implies conditional independence (CI) and an missing edge in the graph of CI under Gaussianity \citep{Lauritzen1996}. For the functional precision matrix, \citet{GGMx} introduced covariate-dependent graphs in $G$ and generalized the concept of CI to CI with covariates (CIx, henceforth). In essence, given a covariate $\bX_i$, the zero precision of $\tilde{\omega}^{j,k}(\bX_i)=0$ implies an missing edge of CIx between nodes $j$ and $k$. Contrarily, when the functional precision is non-zero $\tilde{\omega}^{j,k}(\bX_i)\neq 0$, $Y_j$ and $Y_k$ are conditional dependent with covariates (CDx, henceforth) and an edge exists between nodes $j$ and $k$ given the covariate $\bX_i$. By modeling the functional precision matrix, CIx defines covariate-specific graphs that vary based on different covariates. 


\subsection{Robust Graphical Regression via Random Transformation} \label{sec:rBGR_multiNorm}
In practice, normal assumption does not always hold (as shown in Figure \ref{fig:RPPA_HScore}). Violation of the normal assumption results in the failure of modeling graphs through normal precision matrices and motivates new modeling strategies \citep{tlasso, CSI_Anindya}. In this paper, we adapt the random transformation approach \citep{CSI_Anindya} that allows for various non-normal distributions with different tail behaviors. We focus on continuous distributions with heavy tails as observed in our motivating data. To this end, let $0<d_j<\infty$ for $j=1,\ldots,p$ be independent positive random scales and have distribution as $d_{j} \sim p_j$ with $\int dp(d_{j})<\infty$ almost surely.
Let $\Db_i=\text{diag}(1/d_{i1},\ldots,1/d_{ip})$ be a diagonal matrix for subject $i$. Given random scales 
$d_{ij}, j=1,\ldots,p$ and the covariates $\bX_i$, we assume the distribution of $\Db_i\bY_i$ conditional on $\Db_i$ and $\bX_i$ follows a multivariate distribution,
\begin{align}\label{eq:RandFac_MultiNorm}
    \Db_i\bY_i = \left[\frac{Y_{i1}}{d_{i1}},\ldots,\frac{Y_{ip}}{d_{ip}}\right]^\transp \sim \bN_p(\mathbf{0},\bOmega^{-1}(\bX_i)), \textrm{ for } i=1,\ldots,n,
\end{align}
where $\bOmega(\bX_i)=\{\omega^{j,k}(\bX_i)\}_{p\times p}, j,k\in V$ is the functional precision matrix that characterizes the graph with the covariates $\bX_i$. \\
The model in \eqref{eq:RandFac_MultiNorm} generalizes several existing approaches: (i) Equation \eqref{eq:Gau_MultiNorm} is a special case of Equation \eqref{eq:RandFac_MultiNorm} with $d_{ij}$ as a degenerated distribution of a point mass at one; (ii) when $d_1=\ldots=d_p=\tau$ with $\tau^2$ following an inverse gamma distribution, Equation \eqref{eq:RandFac_MultiNorm} reduced to a multivariate t-distribution on $\bY$ as used by \citet{robustTGraph}, and (iii) for general $d_{ij}$, \eqref{eq:RandFac_MultiNorm} establishes a rich family of Gaussian scale mixtures for the marginal distribution of $Y_j$ with the density $p(Y_j) = \int (2\pi d_j)^{-1/2}\exp\{-y_j^2/(2d_j)\}\textrm{d}p(d_j)$. 

\begin{figure}
    \centering
    \includegraphics[width=\textwidth]{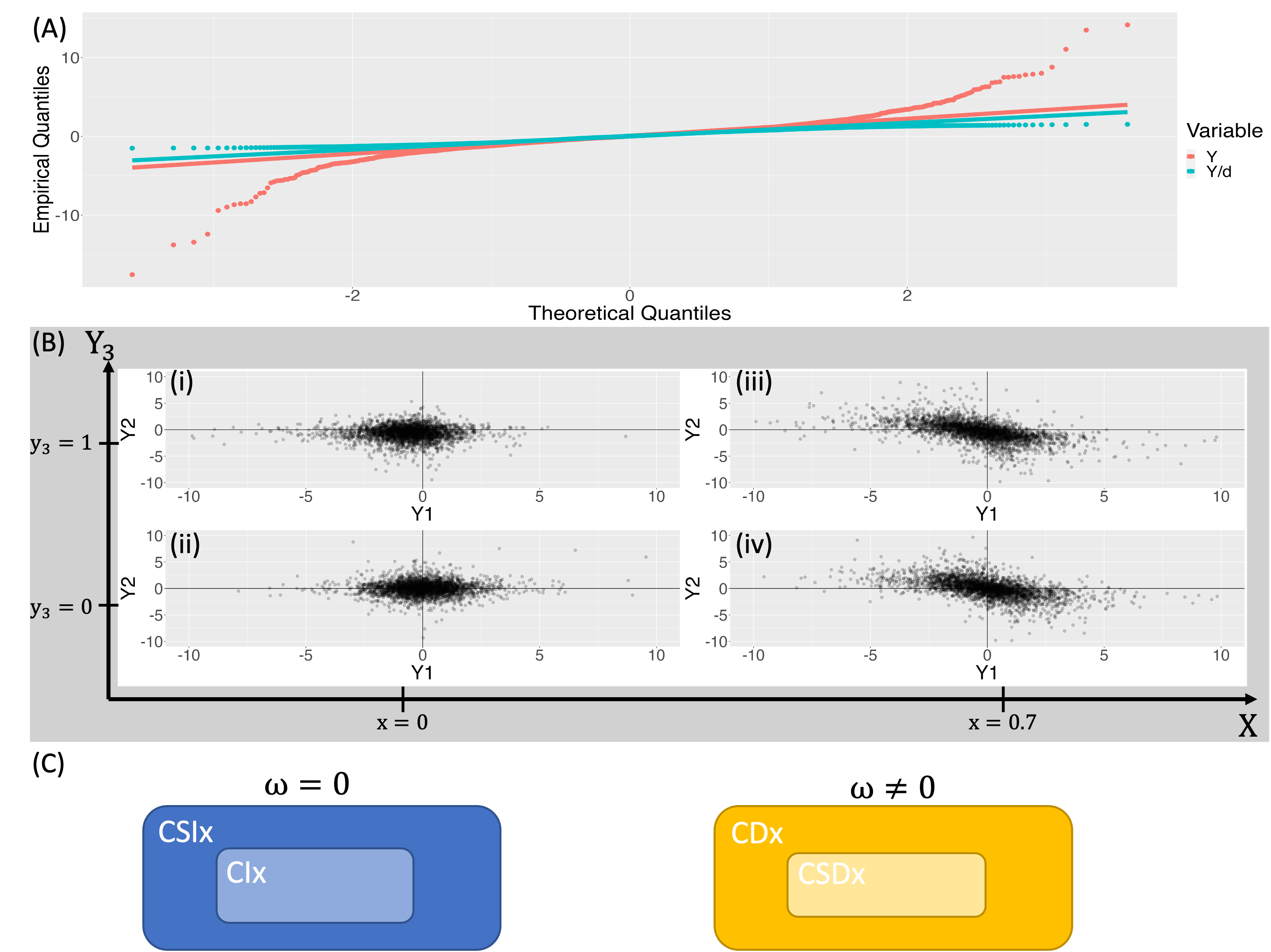}
    \caption{Robustification of non-normal distribution with random scales and the visualization of CSIx and CSDx. Panel (A) shows the qq-plot to illustrate that random scale $d$ accommodates the non-normal distribution $Y$ with $Y/d$ following the normal distribution. Panel (B) demonstrates CSIx (Case (i) and (ii)) and CSDx (Case (iii) and (iv)) of $Y_1$ and $Y_2$ with the partial correlation $\omega^{1,2}(X_i)=X_i$ conditioning on $Y_3$. Cases (i) and (ii) represent two examples of CSIx with zero precision of $X_i=0$ given $Y_3=1$ and $0$. Cases (iii) and (iv) demonstrate the cases of CSDx with non-zero precision of $X_i=0.7$ given $Y_3=1$ and $0$. Panel (B) is centered on the values between $[-10, 10]$. Panel (C) shows the nested relationship between CSIx and CIx (top) and CSDx and CDx (bottom). See more details in Section \ref{sec:char_CSIx}.}
    \label{fig:CSI_Vis}
\end{figure}

The introduction of random scales in Equation \eqref{eq:RandFac_MultiNorm} allows us to construct various marginal distribution of $Y_j$ with different tail behaviors.
Specifically, by matching tail behaviors of random scales and the target distribution, random scales allow us to model different marginal distributions that the data might exhibit. For example, letting $Y_j$ decay polynomially, $Y_j/d_j$ follows a normal distribution if the random scale $d_j$ also has a polynomial tail.  In Figure \ref{fig:CSI_Vis}, Panel (A) shows that the target distribution of $Y$ with a polynomial tail deviates from the normal distribution but with the introduction of random scales the distribution of $Y/d$ is normally distributed. Similar idea can be used for target distribution with other tail behaviours e.g. exponential tails. While the random scales robustify the model to accommodate non-normality, the resulting functional precision matrix $\bOmega(\bX_i)$, however, requires careful characterization and interpretation.

\vspace{-3.5mm}
\vspace{-3.5mm}


\subsection{Characterization of Functional Precision Matrix}  \label{sec:char_CSIx}
The functional precision matrix in \eqref{eq:RandFac_MultiNorm} determines the graphical dependence as a function of covariates, but the random (marginal) scale changes the standard conditional independence interpretations in the resulting precision matrix, which requires a new characterization. \citet{CSI_Anindya} introduced the concept of conditional sign independence (CSI) in non-normal graphs that is defined as follows. Consider two protein expression of interest as random variables $Y_1$ and $Y_2$ with the expression data from the rest of the proteins denoted by a random vector $\bY_3$. Given $\bY_3$, $Y_1$ and $Y_2$ are conditional sign independence (CSI) if $\cP(Y_2>0\mid Y_1,\bY_3) = \cP(Y_2>0\mid \bY_3)$ and $\cP(Y_1>0\mid Y_2,\bY_3) = \cP(Y_1>0\mid \bY_3)$. Otherwise, $Y_1$ and $Y_2$ are conditional sign dependent (CSD) given $\bY_3$. The CSI of $Y_1$ and $Y_2$ implies that the information of $Y_1$ does not affect the sign of $Y_2$ given $\bY_3$. That is, conditioning on the rest of the protein expression data $\bY_3$, the probability of over- or under-expression for $Y_2$ is independent of the expression level of $Y_1$. 
Under the multivariate distribution of \eqref{eq:RandFac_MultiNorm} with a constant precision matrix $\bOmega(\bX_i)=\bOmega$, zero precision of $\omega^{j,k}=0$ and the CSI of $Y_j$ and $Y_k$ given the rest are equivalent, which can be represented by a missing edge between nodes $j$ and $k$ in an undirected graph  \citep{CSI_Anindya, RCGM}.

In this paper, we generalize the concept of CSI to incorporate covariates and consider subject-specific CSI of two random variables given all the other random variables and a realization of covariates $\bX_i$, as formalized in the following proposition:
\begin{proposition}[Conditional Sign Independence with Covariate (CSIx)]\label{prop:CSIx}
Given random scales $\Db_i=\text{diag}(1/d_{i1},\ldots,1/d_{ip})$ and the covariates $\bX_i$, consider the conditional distribution of $\Db_i\bY_i$ as Equation \eqref{eq:RandFac_MultiNorm} with functional precision matrix $\bOmega(\bX_i)$. If $\omega^{j,k}(\bX_i)=0$, then $Y_{ij}$ and $Y_{ik}$ are CSI. 
Otherwise, when $\omega^{j,k}(\bX_i)\neq0$, then $Y_{ij}$ and $Y_{ik}$ are CSD.
\end{proposition} 
The proof of Proposition \ref{prop:CSIx} follows the fact that $\omega^{j,k}(\bX_i)=0$ implies the CSI of $Y_{ij}$ and $Y_{ik}$ given $\bX_i$, and we call $Y_{ij}$ and $Y_{ik}$ are conditional sign independence with covariates  $\bX_i$ to highlight the role of the covariates in the graph. Otherwise, $Y_{ij}$ and $Y_{ik}$ are called CSDx. See Supplementary Material Section S1 for more details.

\noindent {\it An illustrative example.}~ 
We use a simple low-dimensional example to visually demonstrate and interpret CSIx and CSDx. Following Proposition \ref{prop:CSIx}, we show two examples with a general functional precision matrix $\bOmega(\bX_i)$. Consider a trivariate distribution of \eqref{eq:RandFac_MultiNorm} with unit diagonal elements and $\omega^{1,2}(X_i)=X_i$. We illustrate two scenarios shown in Panel (B) of Figure \ref{fig:CSI_Vis}:
\begin{itemize}[noitemsep,topsep=0pt]
\setlength\itemsep{0em}
    \item When $X_i=0$, we obtain the CSIx of $Y_1$ and $Y_2$ given two different values of $Y_3=0$ (Case (i)) and $1$ (Case (ii)).
    \item When $X_i=0.7$, $Y_1$ and $Y_2$ are CSDx and we observe that the distribution of the sign of $Y_2$ varies based on the value of $Y_1$ (see Case (iii) and (iv)). Specifically, as $Y_1$ increases, $Y_2$ tends to be negative. 
\end{itemize}
By modeling the functional precision matrix, we can build covariate-specific precision matrix that depends on the different realization of the covariates $\bX_i$. Consequently, we can construct a graph of CSI corresponding to the precision matrix and the covariates.



We can now conceptually compare models \eqref{eq:Gau_MultiNorm} and \eqref{eq:RandFac_MultiNorm}. Both models incorporate the covariates in the functional precision matrix, which characterize the covariate-specific graph. However, the interpretation of the graph differs. 
The graph from model \eqref{eq:RandFac_MultiNorm} encodes CSIx whereas the graph from model \eqref{eq:Gau_MultiNorm} encodes CIx. We further visualize the relationship between CSIx and CIx in Panel (C) of Figure \ref{fig:CSI_Vis} and summarize as follows: 
\begin{itemize}[noitemsep,topsep=0pt]
    \item For $\omega=0$, CSIx is a weaker condition than CIx since CSIx only considers the sign while CIx depends on both the sign and the magnitude.  
    \item When $\omega\neq 0$, CSDx is a stronger condition than CDx since CDx allows either magnitude or the sign to be dependent while CSDx only focuses on the sign. 
\end{itemize}




\section{Priors and Estimation}\label{sec:estimation}
The functional precision matrix $\bOmega(\bX_i)$ lives in a high-dimensional space. For example, the PPI network for ovarian cancer from our application considers $\mathcal{O}(n\times p\times(p-1)/2)=197,620$ possible edges -- which makes joint estimation difficult if not untenable, especially since variability across each subject $i$ is allowed.  Hence, we employ a neighborhood selection procedure \citep{DAG_NN} to estimate the functional precision matrix that has been used in several graphical modeling approaches \citep[e.g.][]{BGR_Normal,RegGMM}.  This procedure offers three main benefits: (i) tractable estimation, (ii) reduced computation burden, and (iii) flexible prior elicitation. Specifically, we regresses one node $Y_j$ on the rest nodes $Y_k,\; k\neq j$ and build the graph based on zero coefficients (Section \ref{sec:rBGR_Reg} and \ref{sec:RegCoef}). This use of neighborhood selection, which employs conditional estimation as opposed to joint estimation in \eqref{eq:RandFac_MultiNorm} effectively reduce the number of parameters (edges) to $\mathcal{O}(q\times p\times(p-1)/2)=3,280$ -- according a 100-fold reduction. Additionally, the effective number of edges can be further reduced by different model specification such as a thresholding mechanism (Section \ref{sec:CSIF}) and different priors such as spike-and-slab (Section \ref{sec:prior}).


\subsection{Regression-based Approach for Functional Precision Matrix Estimation}\label{sec:rBGR_Reg}
The rBGR model leverages a regression-based framework on model \eqref{eq:RandFac_MultiNorm} to relate the regression coefficients and precision matrix. Given random scales $\bD_i$, we regress one variable on all other variables and relates the partial correlation with regression coefficients. Zero coefficients is then equivalent to zero partial correlations \citep{DAG_NN}. Specifically, we construct node-specific regressions as:
\begin{align}
    \label{eq:rBGR_mod}\frac{Y_{ij}}{d_{ij}} = \sum_{k\neq j}^p \beta_{j,k}({\bf X_i}) \frac{Y_{ik}}{d_{ik}} + \epsilon_{ij},
\end{align}
where $\epsilon_{ij}\sim N(0,1/\omega^{j,j}(\bX_i))$ and the functional coefficient $\beta_{j,k}({\bX_i})=-\frac{\omega^{j,k}(\bX_i)}{\omega^{j,j}(\bX_i)}$. Under this specification, $\beta_{j,k}(\bX_i)=0$ if and only if $\omega^{j,k}(\bX_i)=0$, which enables the functional coefficients to characterize the covariate-specific graphs. However, the interpretation of the coefficients changes from the standard Gaussian graphical models due to the introduction of the random scales, which is detailed in the next subsection.

\subsection{Graph Construction through Regression Coefficients}\label{sec:RegCoef}
We build graphs with a missing edge between nodes $j$ and $k$ when $Y_j$ and $Y_k$ are CSIx given the remaining variables and the covariates $\bX_i$. Consider $\bY_i$ and $\bX_i$ with the regression \eqref{eq:rBGR_mod}.  We call $\beta_{j,k}(\bX_i)$ the conditional sign independence function (CSIF) because zero CSIF $\beta_{j,k}({\bf X}_i)=0$ implies that $Y_j$ and $Y_k$ are CSIx given all the other nodes $\bY_{-\{j,k\}}$ and covariates $\bX_i$, as formally characterized in the following proposition. 
\begin{proposition}\label{Prop_FuncCSI}
    Consider model \eqref{eq:rBGR_mod}. If $\beta_{j,k}({\bf X}_i)=0$, then $\cP(Y_j>0\mid Y_{k}, \bY_{-\{j,k\}},{\bf X}_i) = \cP(Y_j>0\mid \bY_{-\{j,k\}},{\bf X}_i)$ and $\cP(Y_k>0\mid Y_{j}, \bY_{-\{j,k\}},{\bf X}_i) = \cP(Y_k>0\mid \bY_{-\{j,k\}},{\bf X}_i)$.
\end{proposition}
We sketch the proof and leave the details in Supplementary Section S1. The proof follows from the fact that the CSIF $\beta_{j,k}(\bX_i)=-\frac{\omega^{j,k}(\bX_i)}{\omega^{j,j}(\bX_i)}$ is related to the partial correlation, and a zero partial correlation is equivalent to a zero precision of $\omega^{j,k}(\bX)=0$, which ensures the CSIx between $Y_j$ and $Y_k$ (see the example in Section \ref{sec:char_CSIx}). Therefore, zero CSIF indicates the CSIx between $Y_j$ and $Y_k$ given the remaining response variables $\bY_{-\{j,k\}}$ and covariates $\bX_i$. In this paper, we further assume $\omega^{j,j}$ in CSIF to be scalar, $\beta_{j,k}(\bX_i)=-\frac{\omega^{j,k}(\bX_i)}{\omega^{j,j}}$, for ease of computation. Note that our main interest is edge selection and the CSIF is zero if and only $\omega^{j,k}(\bX_i)=0$, which is unrelated to $\omega^{j,j}$.

\subsection{Modeling the Conditional Sign Independence Function}\label{sec:CSIF}
Proposition \ref{Prop_FuncCSI} transforms the problem  of robust graph construction to a more tractable regression coefficient selection (i.e., selecting which part of CSIF is exactly zero). Therefore, modeling the CSIF is crucial to the graph estimation. To this end, we parameterize the CSIF as a product of two components:
\begin{align}
    \beta_{j,k}(\bX_i) = \underbrace{\theta_{j,k}(\bX_i)}_{\textrm{Covariate function}} \underbrace{\mathbb{I}(\mid\theta_{j,k}(\bX_i)\mid>t_{j})}_{\textrm{Thresholding function}}.
\end{align}
We elaborate the role and justification of each component below.

\noindent {\it Covariate functions} [$\theta_{\bullet}(\bullet)$].~
For simplicity, we consider only the linear effects of covariates $\bX_i$, $\theta_{j,k}(\bX_i)=\sum_{h=1}^q \alpha_{j,k,h}X_{ih}$, where $\alpha_{j,k,h}$ represents the coefficients for the $h$-th covariate. The covariate function allows similar edge sets for individuals with a similar level of $\bX_i$ and varies the graph thus borrowing strength. If desired, it is relatively straightforward to extend it to nonlinear effects with e.g., using basis expansion techniques such as splines. 

\noindent  {\it Thresholding functions} [$\mathbb{I}(\mid\theta_{\bullet}(\bX)\mid>t_{\bullet})$].~  
The edge thresholding mechanism is desired to achieve sparse graphs in rBGR due to the large number of parameters and according multiplicity adjustments. For example, the ovarian PPI network in our application requires $qp(p-1)/2=3,280$ parameters and results in a dense graph with inefficient inference. To solve the problem, we truncate edges with small magnitudes with an indicator function $\mathbb{I}(\mid\theta_{j,k}(\bX)\mid>t_{j})$, where $t_{j}$ is the threshold parameter specific to the node $j$. An edge is shrunk to zero and removed when the  magnitude is smaller than the threshold parameter, resulting in a sparse graph. One might consider threshold parameter as $t_{j,k}$. However, $t_{j,k}$ is not fully identifiable when $\alpha_{j,k,h}=0$ for all $h=1,\ldots,q$ since when $\theta_{j,k}(\bX_i)=0$, the value of $t_{j,k}$ can be arbitrary. To alleviate the problem, we assume $t_{j,k}=t_j$ to improve the identifiability as long as one of of $\theta_{j,k}\neq 0$.

\subsection{Prior Specification}\label{sec:prior}
To complete the model specification, rBGR contains three parameters: (a) random scales $d_{j}$, (b) threshold parameter $t_{j}$, and (c) covariate coefficients $\alpha_{j,k,h}$. Specifically, we assign priors as follows:
\begin{align*}
    d_j \sim (1-\pi_j) \delta_1(d_j) + \pi_j p_j(d_j); \thickspace t_j\sim \textrm{Unif}(0,t_{\max})\\
    \alpha_{j,k,h}\sim \gamma_{j,k,h}N(0,\nu_{j,k,h}) + (1-\gamma_{j,k,h})N(0,v_0\nu_{j,k,h}),
\end{align*}
where $v_0$ and $t_{\max}$ are pre-specified hyperparameters, $\pi_j$ models the degree of non-normality with beta prior as $\pi_j\sim Beta(a_\pi, b_\pi)$, $p_j$ is a function to accommodate the non-normality, $\gamma_{j,k,h}$ is a binary variable with Bernoulli prior as $\gamma_{j,k,h}\sim Ber(\rho_j)$, and $\nu_{j,k,h}$ decides the variance of $\alpha_{j,k,h}$ with a inverse Gamma prior of $\nu_{j,k,h}\sim InvGa(a_\nu,b_\nu)$. Specifically, when $d_j=1$, $Y_j$ is normally distributed. When $d_j \sim p_j$, $Y_j$ follows a non-normal distribution. We match tail behavior of $p_j$ and the marginal distribution of $Y_j$, and allow each marginal distribution $Y_j$ to have different level of non-normality by specific $\pi_j$. For the current model, we focus on the $Y_j$ with polynomial decay as illustrated by the motivating data in Figure \ref{fig:RPPA_HScore} and assign a inverse Gamma distribution on $p_j(d_j^2) \sim InvGa(a_{d},b_{d})$. For threshold parameter $t_j$, we assign a uniform prior on $t_{j}$ to model the thresholding mechanism and control the graph sparsity. Intuitively, when $t_{j} \rightarrow 0$, no edge is truncated with a fully connected graph. When $t_{j} \rightarrow \infty$, all edges are shrunk to zero with all nodes disconnected. For covariate coefficients $\alpha_{j,k,h}$, we assign a spike-and-slab prior to achieve the covariate sparsity with a small $v_0$ because not all covariates necessarily contribute to the varying structure of our graph.

\section{Posterior Inference}\label{sec:inference}
{\it Gibbs sampler.}
In this Section, we introduce an efficient Gibbs sampler for the proposed rBGR model. Instead of the Metropolis-Hastings algorithm, we implement a Gibbs sampler except for the random scales, which largely improves the computation and convergence compared to \cite{BGR_Normal}. Recently, \citet{ThrhdMixDist} derived a closed-form of the conditional distributions for Gibbs sampler by formulating the thresholded coefficients as mixture distributions. Specifically, if we view the distribution with one component as a special case of mixture distribution, the mixture distribution from the thresholded coefficient then can achieve conjugacy. We derive the full conditional distributions for covariate coefficients $\alpha_{j,k,h}$ and the threshold parameter $t_j$, which are the mixture of truncated normal and the mixture of uniform distribution, respectively. By assigning normal priors on covariate coefficients and a uniform prior on threshold parameter, we obtain conjugacy on all thresholded parameters. We further use the parameter expansion technique \citep{BGR_Normal,doi:10.1080/01621459.2012.737742} on covariate coefficients to improve the mixing of MCMC. We implement the Metropolis-Hasting algorithm for the random scales. The full details of the full conditional distributions and sampling algorithm are provided in Supplementary Materials (Section S2.2).  
\noindent {\it Covariate and edge selection.} 
The estimated coefficients from rBGR of \eqref{eq:rBGR_mod} do not guarantee symmetry required for undirected graphs. Also, due to the introduction of random scales with CSIx characterization, we focus on the sign of the edge. We describe algorithms to symmetrize the estimated covariate coefficients $\hat{\alpha}_{j,k,h}$ and the sign of graph edges of $\hat{\beta}_{j,k}(\bX_i)$. For covariate coefficients, we compare the posterior inclusion probability (PIP) of asymmetric coefficients from two directions ($\hat{\alpha}_{j,k,h}$ and $\hat{\alpha}_{k,j,h}$) and assign the symmetrized coefficients as the directed coefficient with a smaller PIP. Given a cutoff $c_0$, the rule above requires both directions of coefficients to have PIPs bigger than $c_0$ encouraging a parsimonious network. For the edge, we symmetrize the edge based on the edge posterior probability (ePP). Specifically, we symmetrize ePP by taking the maximum of two asymmetric ePP. Given a cutoff $c_1$, we call an undirected edge if at least one of the directed ePPs is bigger than $c_1$. We then decide the sign of the edge by comparing the posterior probability of positive and negative for the chosen direction. The full details are provided in Supplementary Materials (Section S2.3). 

\section{Simulation Studies}\label{sec:simulation}
We empirically demonstrate the performance of rBGR under a variety of non-normal contaminations and against other competing models in terms of edge and covariate selection. To the best of our knowledge, no other existing method estimates covariate-specific graphs for non-normal data. Therefore, we compare rBGR to two models that estimate the covariate-specific graph without addressing the violation of normality assumption. Specifically, we consider Bayesian graphical regression (BGR) \citep{BGR_Normal} and the Gaussian graphical model regression (RegGMM) \citep{RegGMM} representative of a fully Bayesian and a frequentist penalization-based models for the covariate-specific graph under normal assumption, respectively. For RegGMM, we run the algorithm with various tuning parameters to obtain the probability of the signs of edges and covariate coefficients and select the optimal tuning parameter by cross-validation using their default algorithm. For rBGR and BGR, we symmetrize the graph mentioned in Section \ref{sec:inference} and set $c_0=c_1=0.5$. We run $10,000$ and $30,000$ iterations and discard the first $90\%$ iterations for rBGR and BGR, respectively. 

{\it Data generating mechanism.} 
We generate the observed non-normal data by multiplying the random scale to the latent normal data $\bY_i^*=\left[Y_{i1}^*,\ldots,Y_{ip}^*\right]^\transp$ that follows an multivariate normal distribution with a functional precision matrix that represents the undirected graph. Specifically, we generate the covariates $\bX_i \overset{iid}{\sim} U(-1,1)$ and latent data $\bY_i^* \overset{iid}{\sim} \bN_p(\mathbf{0},\bOmega^{-1}(\bX_i))$ with the true precision matrix $\bOmega(\bX_i)$. For $\bOmega(\bX_i)$, we assign unit diagonal elements and randomly pick $2\%$ of the off-diagonal to be non-zero. We let the non-zero precision depend on the covariates linearly and truncate the precision with a magnitude smaller than $0.15$. We obtain the random scales from a mixture distribution of the point mass at one and a inverse gamma distribution and assign three different levels non-normal contamination: $\pi\in\{0,0.5,0.8\}$. We multiply the random scales to generate the observed data of $[Y_{i1},\ldots, Y_{ip}]=[Y_{i1}^*d_{i1},\ldots,Y_{ip}^*d_{ip}]$. For all simulations, we set the sample size and the dimensions of $\bY_i$ and $\bX_i$ as $(n,p,q)=(250,50,3)$ based on our real data case-studies. We show the results for $50$ independent replicates.

{\it Performance metrics.}
We evaluate the graph recovery through the edge and covariate selection. For covariate selection, we report the true positive rate (TPR), true false rate (TFR), and Matthew’s correlation coefficient (MCC) with the cut-off for PIP at $c_0=0.5$. We also report the area under the ROC curve (AUC) and partial area under ROC curve (pAUC) between specificity ranging from 0.8 to 1. For edge selection, we use AUC and three metrics of TPR, TNR and MCC with the cut-off for ePP at $c_1=0.5$. We further investigate the sign consistency by examining the agreement between the posterior probability for the signs of CSIF $\textrm{sgn}(\hat{\beta}_{j,k}(\bX_i))$ and the true signs of $\textrm{sgn}(\beta_{j,k}(\bX_i))$. Specifically, we exclude the zero CISF and focus on the subset of the data with both true and estimated non-zero CSIF to restrict the problem as two-class classification (positive versus negative). We assess the sign consistency by MCC (referred to as sign-MCC).

\begin{figure}
    \centering
    \includegraphics[width=\textwidth]{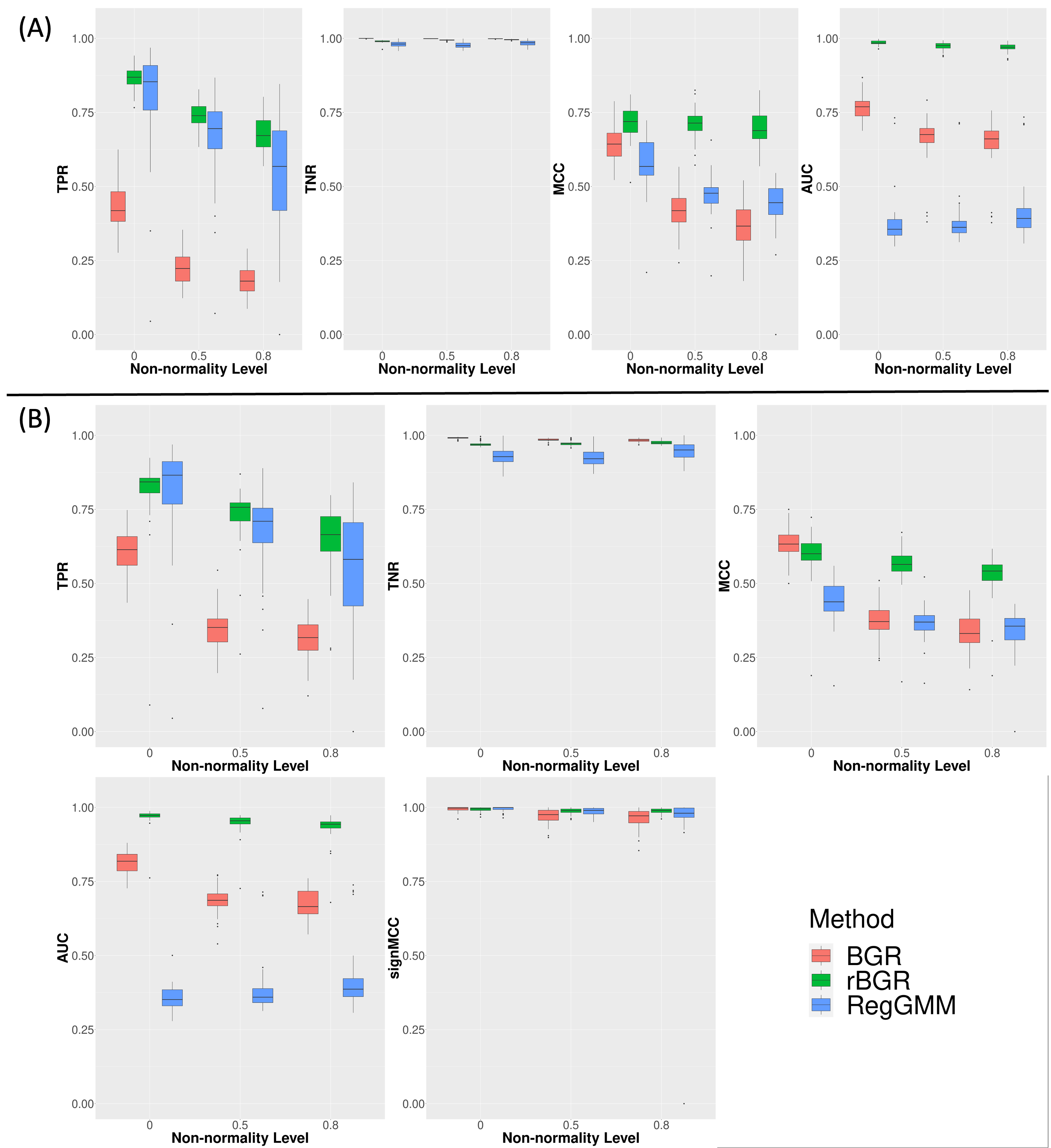}
    \caption{Simulation results: graph recovery for BGR (red), rBGR (green) and RegGMM (blue) under different levels of non-normality in terms of (A) covariates selection (top row) and (B) edge selection (bottom two rows). Panel (A) shows the covariate selection through four metrics (from left to right: TPR, TNR, MCC and AUC) that are measured under three different levels of non-normality. Panel (B) demonstrates the edge selection by four criteria (from upper left to lower right: TPR, TNR, MCC, AUC) and the sign consistency by sign-MCC (lower left) for non-zero edges. All values for TPR, TNR and MCC are measured at a cut-off at $c_0=c_1=0.5$.}
    \label{fig:Sim}
\end{figure}

\noindent {\it Simulation results.} 
Panel (A) of Figure \ref{fig:Sim} shows the simulation results for covariate selection. We observe that rBGR outperforms BGR and RegGMM across all non-normality levels, as indicated by higher MCC and AUC. The difference of MCC and AUC between rBGR and the other competing methods increases when the non-normality contamination level increases, which is expected. For TNR, rBGR performs slightly worse than BGR but better than RegGMM across all non-normality levels. However, all three methods select correct covariates ($>93\%$) with small difference ($<5\%$) in terms of TNR. For TPR, rBGR outperforms BGR under all levels of non-normality and the advantage of rBGR becomes more prominent as the non-normality increases. Compared to RegGMM, rBGR's performance is comparable under normal distribution in TPR, but rBGR is preferred when the level of non-normality increases. 
Overall, modeling the non-normality from random scales in rBGR is favored compared to models without random scales in terms of covariate selection.

We show the graph recovery for the edge selection in Panel (B) of Figure \ref{fig:Sim}. For edge selection, rBGR outperforms BGR and RegGMM in AUC, and the advantage of rBGR increases with a larger discrepancy between rBGR and the competing methods when the non-normality level increases. For MCC, rBGR outperforms RegGMM under all levels of non-normality, but is slightly inferior than BGR under the normal distribution. However, rBGR is favored when the non-normality level increases. For TPR, rBGR is better than BGR under all levels of non-normality, and slightly worse than RegGMM under normal assumption. However, when non-normality increases, rBGR starts to surpass the RegGMM. Both TNR and sign-MCC show excellent selection performance ($>95\%$) for all three methods, with minimal differences ($<5\%$) across the three non-normality levels. 

\underline{In summary}, modeling the non-normality through random scales in rBGR result in equivalent (under normal distribution) or better performances in all metric for edge selection compared to the other methods.

\vspace{-3.5mm}
\vspace{-3.5mm}


\noindent {\it Additional simulations and model evaluations.} 
We provide additional simulation of data generating mechanism and model evaluation results for (i) convergence of the algorithm, and (ii) different cut-off of $c_0$ and $c_1$ controlling for false discovery rates -- which are summarized in Supplementary Material Section S3. Overall, rBGR generates equivalent or better performances compared to other methods given cut-offs based on different criteria.

\section{Analyses of Proteomic Networks under Immunogenic Heterogeneity}\label{sec:realData}

{\it Key scientific questions and dataset overview.}~
Aberrant protein-protein interactions (PPIs) are associated with various diseases including cancer \citep{pmid32228680}, and immune cells around the tumor can modulate malfunctioning PPIs to influence tumor growth and progression \citep{pmid25838376}. In cancer, cells around the tumor form the tumor microenvironment (TME) that closely interacts with the tumor \citep{pmid18836471}. For example, the dysregulated PPIs in tumor suppress multiple immune cells in TME to escape the detection from immune system \citep{pmid18836471} while immune cells in TME can alter the aberrant PPIs to eliminate cancerous cells \citep{pmid25838376}. 
This demonstrates the connection between the dysregulated PPIs and the TME and shows the importance of immunogenic heterogeneity in tumor behavior. A better understanding of the impact of the immune cells on aberrant PPIs offers a foundational paradigm for potential targeted therapies in cancer \citep{pmid32228680}. 
To this end, our key scientific questions were as follows: (i) identify important PPIs across different cancer types and (ii) discover the effect of immunogenic heterogeneity on aberrant PPIs as potential targets for future investigation.

We exemplify the utility of rBGR, using data from The Cancer Genome Atlas (TCGA) to build patient-specific PPI networks and investigate the impact of immunogenic heterogeneity across two different cancers. Specifically, we used reverse-phase protein array for proteomic data ($\bY$) to build the PPI network of CSIx graph and incorporated the immune cell transcriptome signatures as covariates ($\bX$) as markers of immunogenic heterogeneity. Our analysis focuses on ovarian cancer (OV) and lung adenocarcinoma (LUAD) as representative examples of two different types of cancers that elicit distinct immune responses. OV represents a immunologically ``cold" tumor with a weaker immune response, while LUAD is considered a immunologically ``hot" tumor with a stronger immune response \citep{pmid30610226}. 
We focus on proteins in $12$ important cancer-related pathways \citep{PRECISE} and obtained $p=41$ proteins with $n=241$ and $n=360$ patients for OV and LUAD, respectively. For covariates, we included mRNA-derived immune cell gene signatures and quantified the immune cell abundance corresponding to T cells and two crucial members of myeloid-derived suppressor cells, monocytes and neutrophils, for both OV and LUAD. Both T cells and myeloid-derived suppressor cells are essential in both OV and LUAD since T cells is the main immune component that kills cancer cells  while myeloid-derived suppressor cells regulates T cells \citep{pmid18836471}. We ran rBGR on OV and LUAD with $20,000$ iterations and discarded the first $19,000$ iterations. 
The convergence diagnostics and the details of data preprocessing procedures are provided in Supplementary Material Section S4.1.

\subsection{Population-Level Proteomic Networks}
\begin{sidewaysfigure}
    \centering
    \includegraphics[width=\textwidth]{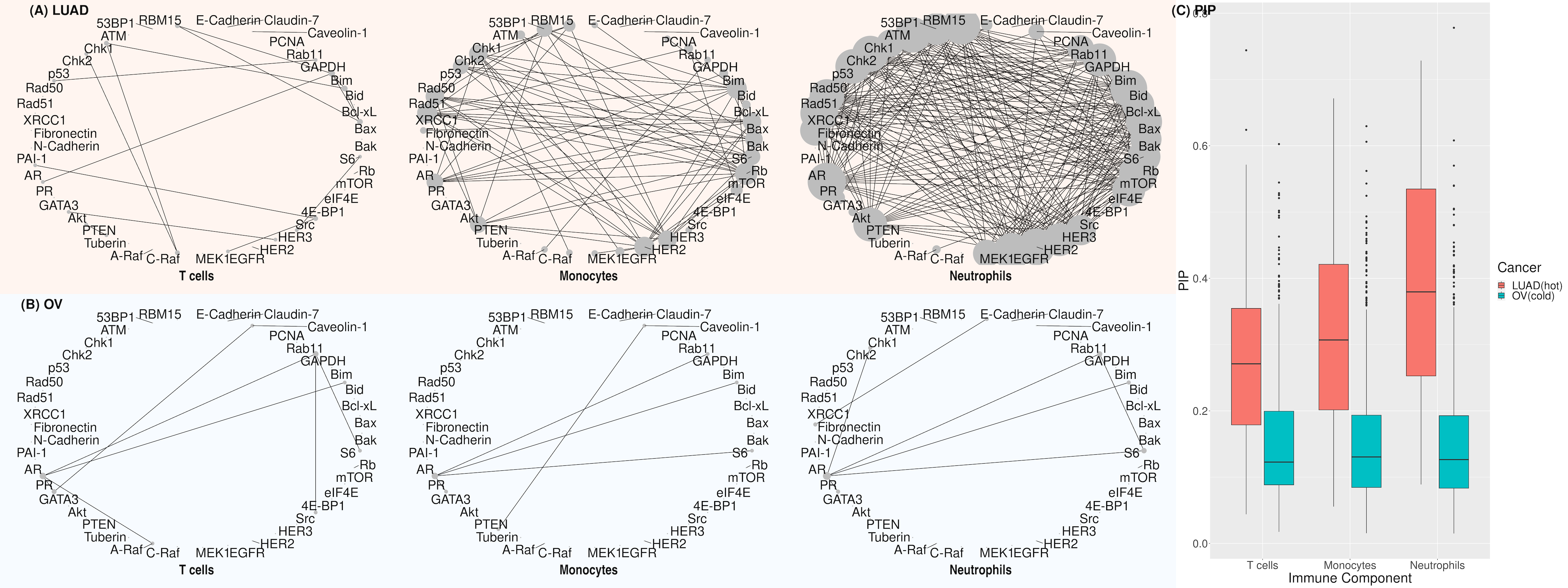}
    \caption{Population-Level Proteomic Networks. The posterior inclusion probability (PIP) (Panel (C)) and the population networks of PPIs with the cut-off at $c_0=0.5$ for LUAD (Panel (A)) and OV (Panel (B)). For each panel of LUAD and OV, PPI networks of specific immune component are shown from the left to right for T cells, monocytes, and neutrophils. The degree of each protein is shown by the node size -- with a bigger node representing a higher degree of connectivity.}
    \label{fig:PopPIP_Net}
\end{sidewaysfigure}

We first focus on the covariate dependent population-level networks for OV and LUAD that are estimated by $\hat{\alpha}_{j,k,h}$. The corresponding networks are shown in in Figure \ref{fig:PopPIP_Net}  (Panels (A) for LUAD and (B) for OV). We observed that the number of edges is much less in OV compared to LUAD for all immune components (T cells: (7, 15), monocytes: (5, 82) and neutrophils: (7, 260) for (OV, LUAD)). This is further evidenced in Panel (C) that shows the distribution of PIPs for OV and LUAD. Interestingly, we observe that the PIPs for LUAD are higher than those for OV for all immune components (median of (OV, LUAD) for T cells: (0.123, 0.271), monocytes: (0.131, 0.307), and neutrophils: (0.127, 0.380)). The higher PIPs in LUAD imply that immune components have a greater impact on PPIs in LUAD compared to OV. This finding is consistent with the existing biology, as LUAD belongs to the immune hot tumors with a stronger immune response \citep{pmid30610226}.
Furthermore, we identify HER2, Rb and Bax as the top three hub proteins with the highest degree in LUAD. 
In LUAD, HER2 mutation is associated inferior survival \citep{pmid28743157}, Bcl-2 family protein including Bax is a prognostic biomarker \citep{pmid28102575}, and Rb mutation predicts poor clinical outcomes \citep{pmid30773851}. For OV, AR protein is identified as a  hub protein with the highest degree (AR: 13 with the rest protein $\leq 10$). Recent evidence supports the critical role of AR for the progression of OV \citep{pmid27741511}. 





Population graphs also confer specific information about the interaction between proteins. For example, we observe an edge between Akt and PTEN with the highest PIP regulated by T cell for LUAD (Panel (A)) suggesting the impact from T cell on the PPI between Akt and PTEN. It is well-known that PTEN down-regulates Akt and the loss of tumor suppressor PTEN often leads to dysregulated PI3K pathway including Akt and the following tumor growth for LUAD \citep{pmid32727102}. 
For OV, despite the smaller number of PPIs, we still identify important PPIs. For example, rBGR suggests a PPI regulated by T cells between Caveolin-1 and PR. In OV, Caveolin-1 is regulated by progesterone, which is mediated by PR, and suggests a consitent result with the estimated PPI between Caveolin-1 and PR \citep{pmid15674352}. 
Overall, our analyses capture important hub proteins and characterize the cancer PPIs, and the results are highly concordant with the existing cancer literature. 

\vspace{-3.5mm}
\vspace{-3.5mm}

\subsection{Patient-Specific Networks: $\beta_{jk}(\bX_i)$}\label{sec:indNet}
We next focus on patient-specific PPI networks to examine the effect of immune component abundance ($\bX_i$) on PPIs. Specifically, we vary one immune component with the rest components fixed at their mean and generate networks of CSIx for different individuals at five percentiles (5th, 25th, 50th, 75th and 95th percentiles) of the varying immune component. We set the cut-off for the ePP at $c_1=0.5$ and show the networks for LUAD in Figure \ref{fig:ePP_ImmuneComp} with the networks for OV in Supplementary Material S4. 


We focus on PPIs of CSDx that show dependency on the abundance of specific immune  component. We present PPIs that change the signs in the 5th and 95th percentiles indicating specific PPIs that are impacted by the immune components, such as Akt-PTEN for T cells, Bid-PCNA for monocytes, and Bax-GATA3 for neutrophils. Interestingly, we discovered that the sign of Akt-PTEN is positively correlated to the T cell abundance. Specifically, when T-cell abundance is higher, Akt-PTEN is positive; vice-versa, Akt-PTEN is negative when T cell is scarce. It is well-established that PTEN suppresses Akt signaling and the loss of PTEN results in the hyper-activation of Akt in cancer cells and the low T cell abundance in lung cancer \citep{pmid32727102}. In addition, we find Bid-PCNA edge is positively correlated with monocytes abundance. It has been shown that PCNA promotes Bid through caspase proteins and is crucial to immune evasion in cancers \citep{pmid34433039}. 
Finally, we discover that Bax-GATA3 edge is positively correlated with neutrophil abundance. Recently, GATA3 has been found to down-regulate BCL-2 \citep{pmid25410484}, which inhibits the Bax protein \citep{pmid9219694}, and neutrophils promotes the Bax to induce the apoptosis \citep{pmid32606746}. 
These findings highlight specific PPIs that are influenced by the abundance of immune components and suggest potential targets for further investigation of immunotherapy.

\vspace{-3.5mm}
\begin{figure}
    \centering
    \includegraphics[width=\textwidth]{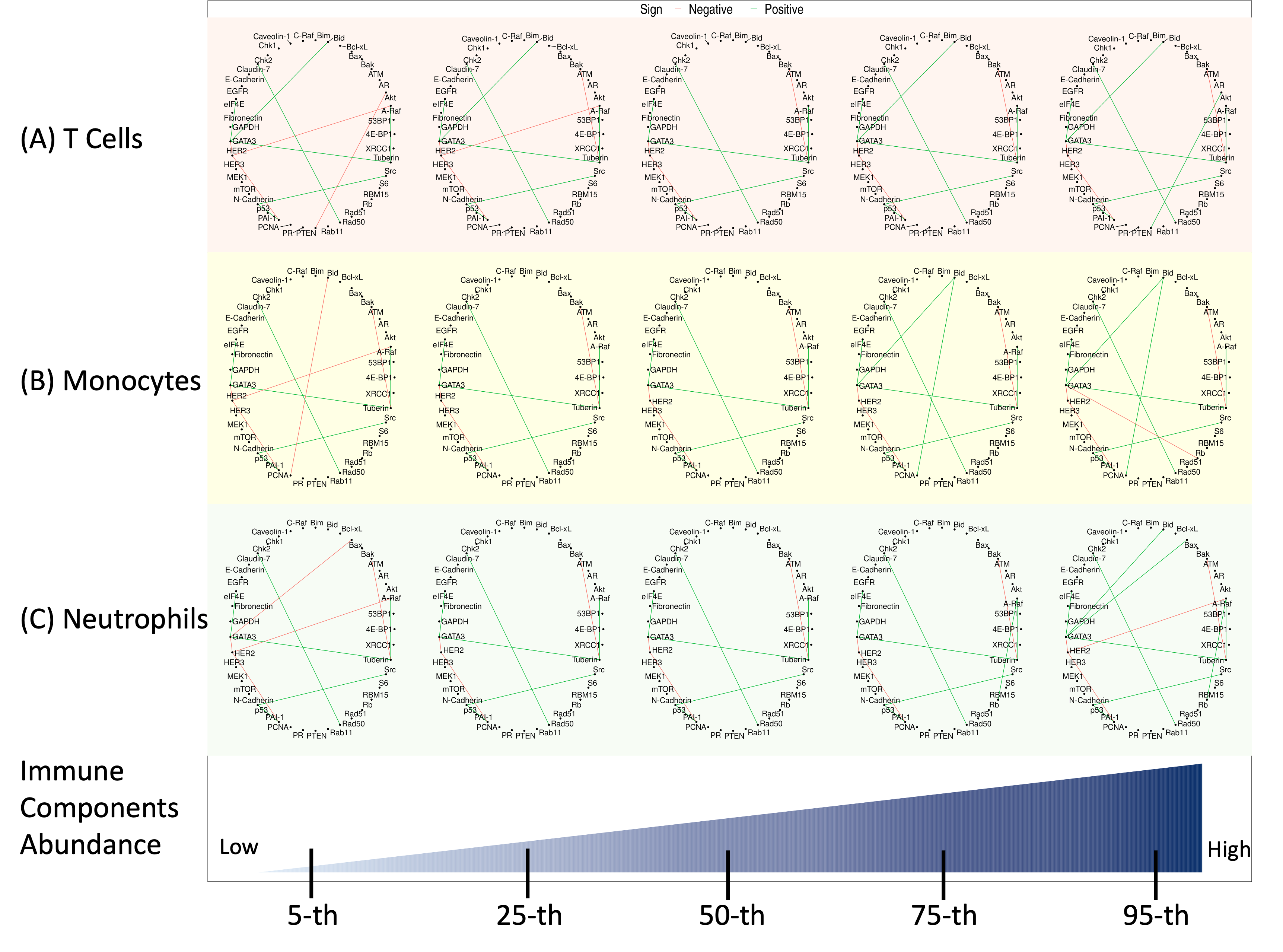}
    \caption{Covariate-specific networks of LUAD under five different percentiles immune component of (A) T cells, (B) monocytes and (C) neutrophils with the rest two components fixed at mean zero. The estimated network for varying immune components are shown from the left to right for 5, 25, 50 ,75, and 95-th percentiles. Edges are identified with signs (green: positive and red: negative) when the ePPs are bigger than $c_1=0.5$.}
    \label{fig:ePP_ImmuneComp}
\end{figure}
\vspace{-3.5mm}

\section{Discussion}\label{sec:Diss}


In this paper, we develop a flexible Bayesian framework called robust Bayesian graphical regression (rBGR) to construct heterogeneous networks that explicitly account for covariate-specific information for non-normally distributed data. By accommodating the non-normal marginal tail behaviors through random scales, we construct covariate-specific graph through graphical regression-based approaches. This framework allows us to explicitly characterize, infer and interpret covariate-specific edge dependencies through conditional sign independence functions. 
%
We also propose an efficient Gibbs sampler for posterior inference. Our simulations demonstrate that rBRG outperforms other existing Gaussian-based methods that construct the covariate-specific graphs under a variety of settings, which display non-normal marginal behavior such as heavy tails or skewness.

We employ rBGR on proteogenomic datasets in two cancers to build patient-specific PPI networks and identify PPIs that are impacted by tumor heterogeneity. Specifically, we quantify the immune cell abundance to elucidate the effects of immunogeneic heterogeneity on aberrant PPI for lung and ovarian cancers that are triggered by different levels of immune responses. Our analyses align with existing biology along three major axes: (i) immune responses, (ii) hub proteins, and (iii) PPIs. For example, higher connections in LUAD are consistent with existing biology since LUAD  belongs to the class of the immunlogically ``hot" tumors. We identify a hub protein of HER2, which is associated with a poor survival in LUAD. Another example is a PPI of Akt-PTEN, which is consistent with the knowledge that PTEN down-regulates Akt. Our study further suggests PPIs that vary with specific immune component. For example, we discover that PPIs of Akt-PTEN, Bid-PCNA, and Bax-GATA3 vary positively with T cells, monocytes and neutrophils, respectively. These findings suggest potential future targets for immunotherapy in lung cancer.

In the current implementation, we consider only linear effect of covariates to reduce the inferential and computation burden. It is possible to include the non-linear functionals through basis expansion techniques such as splines \citep{BGR_Normal} -- however this will increase the computational burden. Another possible extension is other types of graphical dependencies. For example, a chain graph considers ordered multi-level structure via directed and undirected edges \citep[e.g.][]{RCGM}. By introducing the random scales and generalizing the regression coefficients as functional coefficients, the model can include the covariates in the precision matrix to build the subject-specific chain graphs. Another direction could be include discrete nodes and the concept of CSIx can be extended for discrete data \citep{CSI_Anindya}. All these directions are left for future investigations.

{\bf Code and Data Availability.}
We also provide a general purpose code in \texttt{R} that accompanies this manuscript along with all the necessary documentation and datasets required to replicate our results (available at \url{https://github.com/bayesrx/rBGR} with necessary documentation in Supplementary Material Section S5).

\backmatter





%
\bibliographystyle{biom}
\bibliography{rBGR}

\newcommand{\noop}[1]{}
\begin{thebibliography}{}

\bibitem[\protect\citeauthoryear{Airoldi}{Airoldi}{2007}]{Intro_PGM}
Airoldi, E.~M. (2007).
\newblock {{G}etting started in probabilistic graphical models}.
\newblock {\em PLoS Comput Biol} {\bf 3,} e252.

\bibitem[\protect\citeauthoryear{Antonsson, Conti, Ciavatta, Montessuit, Lewis, Martinou, et~al\mbox{.}}{Antonsson et~al.}{1997}]{pmid9219694}
Antonsson, B., Conti, F., Ciavatta, A., Montessuit, S., Lewis, S., Martinou, I., et~al. (1997).
\newblock {{I}nhibition of {B}ax channel-forming activity by {B}cl-2}.
\newblock {\em Science} {\bf 277,} 370--372.

\bibitem[\protect\citeauthoryear{Baladandayuthapani, Talluri, Ji, Coombes, Lu, Hennessy, Davies, and Mallick}{Baladandayuthapani et~al.}{2014}]{pmid26246866}
Baladandayuthapani, V., Talluri, R., Ji, Y., Coombes, K.~R., Lu, Y., Hennessy, B.~T., Davies, M.~A., and Mallick, B.~K. (2014).
\newblock {{B}ayesian sparse graphical models for classification with application to protein expression data}.
\newblock {\em Ann Appl Stat} {\bf 8,} 1443--1468.

\bibitem[\protect\citeauthoryear{Bhadra, Rao, and Baladandayuthapani}{Bhadra et~al.}{2018}]{CSI_Anindya}
Bhadra, A., Rao, A., and Baladandayuthapani, V. (2018).
\newblock {{I}nferring network structure in non-normal and mixed discrete-continuous genomic data}.
\newblock {\em Biometrics} {\bf 74,} 185--195.

\bibitem[\protect\citeauthoryear{Bhateja, Chiu, Wildey, Lipka, Fu, Yang, et~al\mbox{.}}{Bhateja et~al.}{2019}]{pmid30773851}
Bhateja, P., Chiu, M., Wildey, G., Lipka, M.~B., Fu, P., Yang, M. C.~L., et~al. (2019).
\newblock {{R}etinoblastoma mutation predicts poor outcomes in advanced non small cell lung cancer}.
\newblock {\em Cancer Med} {\bf 8,} 1459--1466.

\bibitem[\protect\citeauthoryear{Chakraborty, Baladandayuthapani, Bhadra, and Ha}{Chakraborty et~al.}{2021}]{RCGM}
Chakraborty, M., Baladandayuthapani, V., Bhadra, A., and Ha, M.~J. (2021).
\newblock Bayesian robust learning in chain graph models for integrative pharmacogenomics.

\bibitem[\protect\citeauthoryear{Cheng, Yang, Wang, Leung, and Ma}{Cheng et~al.}{2020}]{pmid32228680}
Cheng, S.~S., Yang, G.~J., Wang, W., Leung, C.~H., and Ma, D.~L. (2020).
\newblock {{T}he design and development of covalent protein-protein interaction inhibitors for cancer treatment}.
\newblock {\em J Hematol Oncol} {\bf 13,} 26.

\bibitem[\protect\citeauthoryear{Chung, Gaynanova, and Ni}{Chung et~al.}{2022}]{chung2022phylogenetically}
Chung, H.~C., Gaynanova, I., and Ni, Y. (2022).
\newblock Phylogenetically informed {B}ayesian truncated copula graphical models for microbial association networks.
\newblock {\em Ann Appl Stat} {\bf 16,} 2437--2457.

\bibitem[\protect\citeauthoryear{Cohen, Ben-Hamo, Gidoni, Yitzhaki, Kozol, Zilberberg, and Efroni}{Cohen et~al.}{2014}]{pmid25410484}
Cohen, H., Ben-Hamo, R., Gidoni, M., Yitzhaki, I., Kozol, R., Zilberberg, A., and Efroni, S. (2014).
\newblock {{S}hift in {G}{A}{T}{A}3 functions, and {G}{A}{T}{A}3 mutations, control progression and clinical presentation in breast cancer}.
\newblock {\em Breast Cancer Res} {\bf 16,} 464.

\bibitem[\protect\citeauthoryear{Conciatori, Bazzichetto, Falcone, Ciuffreda, Ferretti, Vari, et~al\mbox{.}}{Conciatori et~al.}{2020}]{pmid32727102}
Conciatori, F., Bazzichetto, C., Falcone, I., Ciuffreda, L., Ferretti, G., Vari, S., et~al. (2020).
\newblock {{P}{T}{E}{N} function at the interface between cancer and tumor microenvironment: implications for response to immunotherapy}.
\newblock {\em Int J Mol Sci} {\bf 21,}.

\bibitem[\protect\citeauthoryear{Danaher, Wang, and Witten}{Danaher et~al.}{2014}]{pmid24817823}
Danaher, P., Wang, P., and Witten, D.~M. (2014).
\newblock {{T}he joint graphical lasso for inverse covariance estimation across multiple classes}.
\newblock {\em J R Stat Soc Series B Stat Methodol} {\bf 76,} 373--397.

\bibitem[\protect\citeauthoryear{Dobra and Lenkoski}{Dobra and Lenkoski}{2011}]{10.1214/10-AOAS397}
Dobra, A. and Lenkoski, A. (2011).
\newblock {Copula Gaussian graphical models and their application to modeling functional disability data}.
\newblock {\em Ann Appl Stat} {\bf 5,} 969 -- 993.

\bibitem[\protect\citeauthoryear{Finegold and Drton}{Finegold and Drton}{2011}]{tlasso}
Finegold, M. and Drton, M. (2011).
\newblock {Robust graphical modeling of gene networks using classical and alternative t-distributions}.
\newblock {\em Ann Appl Stat} {\bf 5,} 1057 -- 1080.

\bibitem[\protect\citeauthoryear{Finegold and Drton}{Finegold and Drton}{2014}]{robustTGraph}
Finegold, M. and Drton, M. (2014).
\newblock {Robust Bayesian graphical modeling using Dirichlet $t$-distributions}.
\newblock {\em Bayesian Analysis} {\bf 9,} 521 -- 550.

\bibitem[\protect\citeauthoryear{Galon and Bruni}{Galon and Bruni}{2019}]{pmid30610226}
Galon, J. and Bruni, D. (2019).
\newblock {{A}pproaches to treat immune hot, altered and cold tumours with combination immunotherapies}.
\newblock {\em Nat Rev Drug Discov} {\bf 18,} 197--218.

\bibitem[\protect\citeauthoryear{Ha, Banerjee, Akbani, Liang, Mills, Do, and Baladandayuthapani}{Ha et~al.}{2018}]{PRECISE}
Ha, M.~J., Banerjee, S., Akbani, R., Liang, H., Mills, G.~B., Do, K.-A., and Baladandayuthapani, V. (2018).
\newblock Personalized integrated network modeling of the cancer proteome atlas.
\newblock {\em Scientific Reports} {\bf 8,} 14924.

\bibitem[\protect\citeauthoryear{Janku}{Janku}{2014}]{pmid24587830}
Janku, F. (2014).
\newblock {{T}umor heterogeneity in the clinic: is it a real problem?}
\newblock {\em Ther Adv Med Oncol} {\bf 6,} 43--51.

\bibitem[\protect\citeauthoryear{Joyce and Fearon}{Joyce and Fearon}{2015}]{pmid25838376}
Joyce, J.~A. and Fearon, D.~T. (2015).
\newblock {{T} cell exclusion, immune privilege, and the tumor microenvironment}.
\newblock {\em Science} {\bf 348,} 74--80.

\bibitem[\protect\citeauthoryear{Lauritzen}{Lauritzen}{1996}]{Lauritzen1996}
Lauritzen, S.~L. (1996).
\newblock {\em Graphical Models}.
\newblock New York : Oxford University Press.

\bibitem[\protect\citeauthoryear{Li, Li, and Kang}{Li et~al.}{2023}]{ThrhdMixDist}
Li, M., Li, L., and Kang, J. (2023+).
\newblock Bayesian inference of spatially varying correlations via thresholded correlation {G}aussian processes.

\bibitem[\protect\citeauthoryear{Li, Zou, Zhu, Xu, Li, and Zhu}{Li et~al.}{2020}]{pmid32606746}
Li, R., Zou, X., Zhu, T., Xu, H., Li, X., and Zhu, L. (2020).
\newblock Destruction of neutrophil extracellular traps promotes the apoptosis and inhibits the invasion of gastric cancer cells by regulating the expression of bcl-2, bax and nf-κb.
\newblock {\em OncoTargets and Therapy} {\bf 13,} 5271--5281.

\bibitem[\protect\citeauthoryear{Liu, Han, Yuan, Lafferty, and Wasserman}{Liu et~al.}{2012}]{Liu12_NonParaNorm}
Liu, H., Han, F., Yuan, M., Lafferty, J., and Wasserman, L. (2012).
\newblock {High-dimensional semiparametric Gaussian copula graphical models}.
\newblock {\em Ann Stat} {\bf 40,} 2293 -- 2326.

\bibitem[\protect\citeauthoryear{Meinshausen and B{\"u}hlmann}{Meinshausen and B{\"u}hlmann}{2006}]{DAG_NN}
Meinshausen, N. and B{\"u}hlmann, P. (2006).
\newblock {High-dimensional graphs and variable selection with the Lasso}.
\newblock {\em Ann Stat} {\bf 34,} 1436 -- 1462.

\bibitem[\protect\citeauthoryear{Ni, Baladandayuthapani, Vannucci, and Stingo}{Ni et~al.}{2022}]{Ni2022}
Ni, Y., Baladandayuthapani, V., Vannucci, M., and Stingo, F.~C. (2022).
\newblock Bayesian graphical models for modern biological applications.
\newblock {\em Stat Methods {\&} Appl} {\bf 31,} 197--225.

\bibitem[\protect\citeauthoryear{Ni, Stingo, and Baladandayuthapani}{Ni et~al.}{2019}]{BGR_Normal}
Ni, Y., Stingo, F.~C., and Baladandayuthapani, V. (2019).
\newblock Bayesian graphical regression.
\newblock {\em J Am Stat Assoc} {\bf 114,} 184--197.

\bibitem[\protect\citeauthoryear{Ni, Stingo, and Baladandayuthapani}{Ni et~al.}{2022}]{GGMx}
Ni, Y., Stingo, F.~C., and Baladandayuthapani, V. (2022).
\newblock Bayesian covariate-dependent {G}aussian graphical models with varying structure.
\newblock {\em Journal of Machine Learning Research} {\bf 23,} 1--29.

\bibitem[\protect\citeauthoryear{Peng, Wang, Zhou, and Zhu}{Peng et~al.}{2009}]{pmid19881892}
Peng, J., Wang, P., Zhou, N., and Zhu, J. (2009).
\newblock {{P}artial correlation estimation by joint sparse regression models}.
\newblock {\em J Am Stat Assoc} {\bf 104,} 735--746.

\bibitem[\protect\citeauthoryear{Peterson, Stingo, and Vannucci}{Peterson et~al.}{2015}]{pmid26078481}
Peterson, C.~B., Stingo, F.~C., and Vannucci, M. (2015).
\newblock {{B}ayesian inference of multiple {G}aussian graphical models}.
\newblock {\em J Am Stat Assoc} {\bf 110,} 159--174.

\bibitem[\protect\citeauthoryear{Pillai, Behera, Berry, Rossi, Kris, Johnson, Bunn, Ramalingam, and Khuri}{Pillai et~al.}{2017}]{pmid28743157}
Pillai, R.~N., Behera, M., Berry, L.~D., Rossi, M.~R., Kris, M.~G., Johnson, B.~E., Bunn, P.~A., Ramalingam, S.~S., and Khuri, F.~R. (2017).
\newblock {{H}{E}{R}2 mutations in lung adenocarcinomas: {A} report from the {L}ung {C}ancer {M}utation {C}onsortium}.
\newblock {\em Cancer} {\bf 123,} 4099--4105.

\bibitem[\protect\citeauthoryear{Pitt, Chan, and Kohn}{Pitt et~al.}{2006}]{10.2307/20441306}
Pitt, M., Chan, D., and Kohn, R. (2006).
\newblock Efficient {B}ayesian inference for {G}aussian copula regression models.
\newblock {\em Biometrika} {\bf 93,} 537--554.

\bibitem[\protect\citeauthoryear{Scheipl, Fahrmeir, and Kneib}{Scheipl et~al.}{2012}]{doi:10.1080/01621459.2012.737742}
Scheipl, F., Fahrmeir, L., and Kneib, T. (2012).
\newblock Spike-and-slab priors for function selection in structured additive regression models.
\newblock {\em J Am Stat Assoc} {\bf 107,} 1518--1532.

\bibitem[\protect\citeauthoryear{Sun, Sasano, and Gao}{Sun et~al.}{2017}]{pmid28102575}
Sun, P.~L., Sasano, H., and Gao, H. (2017).
\newblock {{B}cl-2 family in non-small cell lung cancer: its prognostic and therapeutic implications}.
\newblock {\em Pathol Int} {\bf 67,} 121--130.

\bibitem[\protect\citeauthoryear{Syed, Mukherjee, Lyons-Weiler, Lau, Mashima, Tsuruo, and Ho}{Syed et~al.}{2005}]{pmid15674352}
Syed, V., Mukherjee, K., Lyons-Weiler, J., Lau, K.~M., Mashima, T., Tsuruo, T., and Ho, S.~M. (2005).
\newblock {{I}dentification of {A}{T}{F}-3, caveolin-1, {D}{L}{C}-1, and {N}{M}23-{H}2 as putative antitumorigenic, progesterone-regulated genes for ovarian cancer cells by gene profiling}.
\newblock {\em Oncogene} {\bf 24,} 1774--1787.

\bibitem[\protect\citeauthoryear{Wang, Lee, Shen, Lin, Wu, Lin, et~al\mbox{.}}{Wang et~al.}{2021}]{pmid34433039}
Wang, Y.~L., Lee, C.~C., Shen, Y.~C., Lin, P.~L., Wu, W.~R., Lin, Y.~Z., et~al. (2021).
\newblock {{E}vading immune surveillance via tyrosine phosphorylation of nuclear {P}{C}{N}{A}}.
\newblock {\em Cell Rep} {\bf 36,} 109537.

\bibitem[\protect\citeauthoryear{Weinstein, Collisson, Mills, Shaw, Ozenberger, Ellrott, et~al\mbox{.}}{Weinstein et~al.}{2013}]{pmid24071849}
Weinstein, J.~N., Collisson, E.~A., Mills, G.~B., Shaw, K.~R., Ozenberger, B.~A., Ellrott, K., et~al. (2013).
\newblock {{T}he {C}ancer {G}enome {A}tlas {P}an-{C}ancer analysis project}.
\newblock {\em Nat Genet} {\bf 45,} 1113--1120.

\bibitem[\protect\citeauthoryear{Whiteside}{Whiteside}{2008}]{pmid18836471}
Whiteside, T.~L. (2008).
\newblock {{T}he tumor microenvironment and its role in promoting tumor growth}.
\newblock {\em Oncogene} {\bf 27,} 5904--5912.

\bibitem[\protect\citeauthoryear{Zhang and Li}{Zhang and Li}{2022}]{RegGMM}
Zhang, J. and Li, Y. (2022).
\newblock High-dimensional {G}aussian graphical regression models with covariates.
\newblock {\em J Am Stat Assoc} {\bf 0,} 1--13.

\bibitem[\protect\citeauthoryear{Zhu, Zhu, Zheng, Hu, Sun, and Zhu}{Zhu et~al.}{2017}]{pmid27741511}
Zhu, H., Zhu, X., Zheng, L., Hu, X., Sun, L., and Zhu, X. (2017).
\newblock {{T}he role of the androgen receptor in ovarian cancer carcinogenesis and its clinical implications}.
\newblock {\em Oncotarget} {\bf 8,} 29395--29405.

\end{thebibliography}


 \newcommand{\noop}[1]{}
\begin{thebibliography}{}

\bibitem[Armstrong, 2014]{pmid24697967}
Armstrong, R.~A. (2014).
\newblock {{W}hen to use the {B}onferroni correction}.
\newblock {\em Ophthalmic Physiol Opt}, 34(5):502--508.

\bibitem[Bhadra et~al., 2018]{CSI_Anindya}
Bhadra, A., Rao, A., and Baladandayuthapani, V. (2018).
\newblock {{I}nferring network structure in non-normal and mixed discrete-continuous genomic data}.
\newblock {\em Biometrics}, 74(1):185--195.

\bibitem[Geweke, 1992]{Geweke92evaluatingthe}
Geweke, J. (1992).
\newblock Evaluating the accuracy of sampling-based approaches to the calculation of posterior moments.
\newblock In {\em In Bayesian Statistics}, pages 169--193. University Press.

\bibitem[Ha et~al., 2018]{PRECISE}
Ha, M.~J., Banerjee, S., Akbani, R., Liang, H., Mills, G.~B., Do, K.-A., and Baladandayuthapani, V. (2018).
\newblock Personalized integrated network modeling of the cancer proteome atlas.
\newblock {\em Scientific Reports}, 8(1):14924.

\bibitem[Lauritzen, 1996]{Lauritzen1996}
Lauritzen, S.~L. (1996).
\newblock {\em Graphical Models}.
\newblock New York : Oxford University Press.

\bibitem[Li et~al., 2023]{ThrhdMixDist}
Li, M., Li, L., and Kang, J. (2023+).
\newblock Bayesian inference of spatially varying correlations via thresholded correlation {G}aussian processes.

\bibitem[Luo et~al., 2021]{pmid34248988}
Luo, X., Xu, J., Yu, J., and Yi, P. (2021).
\newblock {{S}haping immune responses in the tumor microenvironment of ovarian cancer}.
\newblock {\em Front Immunol}, 12:692360.

\bibitem[Martins-Cardoso et~al., 2020]{pmid32545405}
Martins-Cardoso, K., Almeida, V.~H., Bagri, K.~M., Rossi, M. I.~D., Mermelstein, C.~S., nig, S., and Monteiro, R.~Q. (2020).
\newblock {{N}eutrophil extracellular traps ({N}{E}{T}s) promote pro-metastatic phenotype in human breast cancer cells through epithelial-mesenchymal transition}.
\newblock {\em Cancers (Basel)}, 12(6).

\bibitem[Meinshausen and B{\"u}hlmann, 2006]{DAG_NN}
Meinshausen, N. and B{\"u}hlmann, P. (2006).
\newblock {High-dimensional graphs and variable selection with the Lasso}.
\newblock {\em Ann Stat}, 34(3):1436 -- 1462.

\bibitem[Ni et~al., 2019]{BGR_Normal}
Ni, Y., Stingo, F.~C., and Baladandayuthapani, V. (2019).
\newblock Bayesian graphical regression.
\newblock {\em J Am Stat Assoc}, 114(525):184--197.

\bibitem[Nirmal et~al., 2018]{ImSig}
Nirmal, A.~J., Regan, T., Shih, B.~B., Hume, D.~A., Sims, A.~H., and Freeman, T.~C. (2018).
\newblock {{I}mmune cell gene signatures for profiling the microenvironment of solid tumors}.
\newblock {\em Cancer Immunol Res}, 6(11):1388--1400.

\bibitem[Scheipl et~al., 2012]{doi:10.1080/01621459.2012.737742}
Scheipl, F., Fahrmeir, L., and Kneib, T. (2012).
\newblock Spike-and-slab priors for function selection in structured additive regression models.
\newblock {\em J Am Stat Assoc}, 107(500):1518--1532.

\bibitem[Spella and Stathopoulos, 2021]{pmid33494181}
Spella, M. and Stathopoulos, G.~T. (2021).
\newblock {{I}mmune resistance in lung adenocarcinoma}.
\newblock {\em Cancers (Basel)}, 13(3).

\bibitem[Storey and Tibshirani, 2003]{FDR_Paper}
Storey, J.~D. and Tibshirani, R. (2003).
\newblock {Statistical significance for genomewide studies}.
\newblock {\em Proceedings of the National Academy of Sciences of the United States of America}, 100(16):9440--9445.

\bibitem[Wang et~al., 2022]{pmid36008393}
Wang, C., Yu, Q., Song, T., Wang, Z., Song, L., Yang, Y., et~al. (2022).
\newblock {{T}he heterogeneous immune landscape between lung adenocarcinoma and squamous carcinoma revealed by single-cell {R}{N}{A} sequencing}.
\newblock {\em Signal Transduct Target Ther}, 7(1):289.

\bibitem[Yang et~al., 2020]{pmid32850861}
Yang, Y., Yang, Y., Yang, J., Zhao, X., and Wei, X. (2020).
\newblock {{T}umor microenvironment in ovarian cancer: function and therapeutic strategy}.
\newblock {\em Front Cell Dev Biol}, 8:758.

\bibitem[Zhang and Li, 2022]{RegGMM}
Zhang, J. and Li, Y. (2022).
\newblock High-dimensional {G}aussian graphical regression models with covariates.
\newblock {\em J Am Stat Assoc}, 0(0):1--13.

\end{thebibliography}








\label{lastpage}

\end{document}


\setstcolor{red}

\def\spacingset#1{\renewcommand{\baselinestretch}%
{#1}\small\normalsize} \spacingset{1}
\def\0{\mbox{\boldmath{$\mathbf{0}$}}}
\def\1{\mbox{\boldmath{$\mathbf{1}$}}}
\def\bzeta{\mbox{\boldmath$\zeta$}}
\def\bnu{\mbox{\boldmath$\nu$}}
\def\btheta{\mbox{\boldmath$\theta$}}
\def\bTheta{\mbox{\boldmath$\Theta$}}
\def\bmu{\mbox{\boldmath$\mu$}}
\def\bbeta{\mbox{\boldmath$\beta$}}
\def\bchi{\mbox{\boldmath$\chi$}}
\def\boldeta{\mbox{\boldmath$\eta$}}
\def\bzeta{\mbox{\boldmath$\zeta$}}
\def\bepsilon{\mbox{\boldmath$\epsilon$}}
\def\bomega{\mbox{\boldmath$\omega$}}
\def\bOmega{\mbox{\boldmath$\Omega$}}
\def\bgamma{\mbox{\boldmath$\gamma$}}
\def\bGamma{\mbox{\boldmath$\Gamma$}}
\def\bsigma{\mbox{\boldmath$\sigma$}}
\def\bSigma{\mbox{\boldmath$\Sigma$}}
\def\bdelta{\mbox{\boldmath$\delta$}}
\def\bDelta{\mbox{\boldmath$\Delta$}}
\def\blambda{\mbox{\boldmath$\lambda$}}
\def\bLambda{\mbox{\boldmath$\Lambda$}}
\def\calF{\mbox{$\mathcal{F}$}}
\def\calH{\mbox{$\mathcal{H}$}}
\def\calI{\mbox{$\mathcal{I}$}}
\def\calJ{\mbox{$\mathcal{J}$}}
\def\calX{\mbox{$\mathcal{X}$}}
\def\A{\mbox{\boldmath{$\mathbf{A}$}}}
\def\B{\mbox{\boldmath{$\mathbf{B}$}}}
\def\C{\mbox{\boldmath{$\mathbf{C}$}}}
\def\D{\mbox{\boldmath{$\mathbf{D}$}}}
\def\G{\mbox{\boldmath{$\mathbf{G}$}}}
\def\I{\mbox{\boldmath{$\mathbf{I}$}}}
\def\J{\mbox{\boldmath{$\mathbf{J}$}}}
\def\M{\mbox{\boldmath{$\mathbf{M}$}}}
\def\N{\mbox{\boldmath{$\mathbf{N}$}}}
\def\R{\mbox{\boldmath{$\mathbf{R}$}}}
\def\S{\mbox{\boldmath{$\mathbf{S}$}}}
\def\U{\mbox{\boldmath{$\mathbf{U}$}}}
\def\X{\mbox{\boldmath{$\mathbf{X}$}}}
\def\W{\mbox{\boldmath{$\mathbf{W}$}}}
\def\Y{\mbox{\boldmath{$\mathbf{Y}$}}}
\def\Z{\mbox{\boldmath{$\mathbf{Z}$}}}
\def\a{\mbox{\boldmath{$\mathbf{a}$}}}
\def\c{\mbox{\boldmath{$\mathbf{c}$}}}
\def\e{\mbox{\boldmath{$\mathbf{e}$}}}
\def\f{\mbox{\boldmath{$\mathbf{f}$}}}
\def\h{\mbox{\boldmath{$\mathbf{h}$}}}
\def\j{\mbox{\boldmath{$\mathbf{j}$}}}
\def\p{\mbox{\boldmath{$\mathbf{p}$}}}
\def\x{\mbox{\boldmath{$\mathbf{x}$}}}
\def\y{\mbox{\boldmath{$\mathbf{y}$}}}
\def\z{\mbox{\boldmath{$\mathbf{z}$}}}


\if1\blind
{
  \title{\bf Supplementary Material for ``Robust Bayesian Graphical Regression Models for Assessing Tumor Heterogeneity in Proteomic Networks''}
  \author{Tsung-Hung Yao$^{1,*}$, Yang Ni$^{2}$, Anindya Bhadra$^{3}$, Jian Kang$^{1}$\\
		    and Veerabhadran Baladandayuthapani$^{a}$\\~\\
			$^{1}$Department of Biostatistics, University of Michigan\\
			$^{2}$Department of Statistics, Texas A\&M University\\
			$^{3}$Department of Statistics, Purdue University\\
			$^{*}${\small corresponding author. E-mail: yaots@umich.edu}
			}
  \maketitle
} \fi

\if0\blind
{
  \bigskip
  \bigskip
  \begin{center}
    {\Large\bf Supplementary Material for ``Robust Bayesian Graphical Regression Models for Assessing Tumor Heterogeneity in Proteomic Networks''}
\end{center}
  \medskip
} \fi

\bigskip

\spacingset{1.45} 

\tableofcontents

\section{Proofs}\label{supp:PropProof}
We provide a detailed proof for Propositions 1 and 2 in the Main Paper. 

\subsection{Proof of Proposition 1}
\begin{proof}
For the scale case of precision matrix of $\Omega^{j,k}(\bX_i)=\Omega^{j,k}$, it is well-known that the zero precision $\omega^{j,k}=0$ implies the CSI of $Y_{ij}$ and $Y_{ik}$ \citep{CSI_Anindya}. For the functional precision matrix, we incorporate the covariates in the precision by an indicator function as $\mathbb{I}(\omega^{j,k}(\bX_i))=0$. By doing so, the CSI of $Y_{ij}$ and $Y_{ik}$ still holds. 
\end{proof}

\subsection{Proof of Proposition 2}
We proceed this proof through two steps. First, we show the conditional sign independence for the undirected graph and the equivalent graphical regression model without covariates. We then can extend to result to the regression model for conditional sign independence with covariates.

\begin{proof}
We first show the undirected case with scalar coefficients $\beta$. Denote the $\bD=\textrm{diag}(\frac{1}{d_1},\ldots,\frac{1}{d_p})$ a diagonal matrix of dimension $p$ by $p$. Following the assumption of normal conditional distribution of (3), the joint distribution of $\bY\bD=[\frac{Y_1}{d_1},\ldots,\frac{Y_p}{d_p}]$ is a multivariate normal distribution $\bN_p(\bmu,\bSigma)$. From the Proposition (C.5) of \citet{Lauritzen1996}, we can first partition the joint distribution with 
        \begin{align*}
            \bY\bD \mid \bD=\begin{bmatrix}Y_j/d_{j} \\ Y_{j'}/d_{j'} \\ \bY_{V\setminus\{j,j'\}}\bD_{V\setminus\{j,j'\}}
            \end{bmatrix} \sim 
            \bN_p\left(\begin{bmatrix} \mu_j \\ \mu_{j'} \\ \bmu_{V\setminus\{j,j'\}}
            \end{bmatrix}, 
            \begin{bmatrix}
                \kappa_{jj} & \kappa_{jj'} & \bkappa_{j.}^\transp \\ 
                \kappa_{jj'} & \kappa_{jj} & \bkappa_{j'.}^\transp \\
                \bkappa_{j.} & \bkappa_{j'.} & \cK_{V\setminus\{j',j\}}
            \end{bmatrix}^{-1}
            \right)
        \end{align*}
        where $\bSigma=\cK^{-1}$, $\bkappa_{j.}=[\kappa_{jv}]$ and  $\bkappa_{j'.}=[\kappa_{{j'}v}], v\in V\setminus\{j,j'\}$. Thus, the conditional distribution of $\frac{Y_j}{d_j}$ and $\frac{Y_{j'}}{d_{j'}}$ is a bivariate normal distribution:
        \begin{align}\label{eq:biNorm}
            \left.\begin{bmatrix}Y_j/d_j \\ Y_{j'}/d_{j'}\end{bmatrix}\right\vert \bY_{V\setminus\{j,j'\}},\bD  \sim \bN_2\left(\begin{bmatrix}\mu_{jD}\\\mu_{{j'}D}\end{bmatrix}, \cK_{jj'}^{-1}\right),
        \end{align}
	    where $\cK_{j{j'}}=\begin{bmatrix}\kappa_{jj} & \kappa_{j{j'}} \\ \kappa_{j{j'}} & \kappa_{j'j'}\end{bmatrix}$ and 
	    $\begin{bmatrix} \mu_{jD} \\ \mu_{{j'}D}\end{bmatrix} = 
	    \begin{bmatrix} \mu_{j} \\ \mu_{{j'}} \end{bmatrix} - \cK_{j{j'}}^{-1} \begin{bmatrix} \bkappa_{j.}^\transp \\ \bkappa_{j'.}^\transp \end{bmatrix} (\bY_{V\setminus\{j,j'\}}\bD_{V\setminus\{j,j'\}} - \bmu_{V\setminus\{j,j'\}}) = 
	    \begin{bmatrix} \mu_{j} \\ \mu_{j'} \end{bmatrix} - \cK_{jj'}^{-1}
	    \begin{bmatrix}
	    \sum_{v\in V\setminus\{j,j'\}} \kappa_{jv}(Y_v/d_v - \mu_v) \\ \sum_{v\in V\setminus\{j,j'\}} \kappa_{j'v}(Y_v/d_v - \mu_v)
	    \end{bmatrix}$. 
	    
	    Now, we can show the univariate distribution of $\frac{Y_j}{d_j}$:
	    \begin{align*}
	        \left. \frac{Y_j}{d_j} \right\vert \bY_{V\setminus\{j\}},\bD  \sim N\big(\tilde{\mu}_{jD}, \kappa^{-1}_{jj}\big),
	    \end{align*}
	    where $\tilde{\mu}_{jD}=\mu_j - \frac{1}{\kappa_{jj}} \begin{bmatrix}\kappa_{jj'} & \bkappa^\transp_{j.}\end{bmatrix}(\bY_{V\setminus\{j\}}\bD_{V\setminus\{j\}} - \bmu_{V\setminus\{j\}})=
	    \mu_j - \frac{1}{\kappa_{jj}}\sum_{v\in V\setminus\{j\}} \kappa_{jv}(Y_v/d_v - \mu_v)$.
	    When $Y_j/d_j$ and $Y_{j'}/d_{j'}$ are independent, $\kappa_{jj'}=0$ and $\mu_{jD}=\mu_j - \kappa_{jj}^{-1}\sum_{v\in V\setminus\{j,j'\}} \kappa_{jv}(Y_v/d_v - \mu_v) = \tilde{\mu}_{jD}$. Thus,
	    \begin{align}\label{eq:uniNorm}
	        p(Y_j/d_j \vert \bY_{V\setminus\{j,j'\}},\bD) = p(Y_j/d_j \vert \bY_{V\setminus\{j\}},\bD).
	    \end{align}
	    However, the conditional independence does not hold after integrating out the random scaling $\bD$. Specifically, the integration of $\bD$ conditioning on $\bY_{V\setminus\{j,j'\}}$ and $\bY_{V\setminus\{j\}}$ are different. We can see that from the following expectation values.
	    \begin{gather*}
	    \mathbb{E}[Y_j\vert \bY_{V\setminus\{j,{j'}\}}] =\mathbb{E}_{\bD\vert \bY_{V\setminus\{j,{j'}\}}}[\mathbb{E}[Y_j\vert \bY_{V\setminus\{j,{j'}\}}, \bD]] = \mathbb{E}_{\bD\vert \bY_{V\setminus\{j,{j'}\}}}[d_j\mu_{jD}]\\
	    \mathbb{E}[Y_j\vert \bY_{V\setminus\{j\}}] =\mathbb{E}_{\bD\vert \bY_{V\setminus\{j\}}}\mathbb{E}[[Y_j\vert \bY_{V\setminus\{j\}}, \bD]]= \mathbb{E}_{\bD\vert \bY_{V\setminus\{j\}}}[d_j\mu_{jD}]
	    \end{gather*}
	    Since the conditional distributions of $\bD\vert \bY_{V\setminus\{j\}}$ and $\bD\vert \bY_{V\setminus\{j,j'\}}$ are not equal, the expectation values are different. Of note, the conditional sign independence still hold due to following equations:
	    \begin{align*}
	        \mathbb{P}(Y_j<0\vert \bY_{V\setminus\{j,j'\}}) &= \mathbb{E}_{\bD\vert \bY_{V\setminus\{j,j'\}}}\left[\mathbb{P}(Y_j<0\vert \bY_{V\setminus\{j,j'\}},\bD)\right]\\
	        &= \mathbb{E}_{\bD\vert \bY_{V\setminus\{j,j'\}}}\left[\mathbb{P}(Y_j/d_j<0\vert \bY_{V\setminus\{j,{j'}\}},\bD)\right] \\
	        &= \mathbb{E}_{\bD\vert \bY_{V\setminus\{j,j'\}}}\left[\mathbb{P}(\kappa_{jj}^{1/2}(Y_j/d_j - \mu_{jD})< -\kappa_{jj}^{1/2}\mu_{jD}\vert \bY_{V\setminus\{j,j'\}},\bD)\right]\\
	        &= \mathbb{E}_{\bD_{V\setminus\{j,j'\}}\vert \bY_{V\setminus\{j,j'\}}}\left[\Phi(-\kappa_{jj}^{1/2}\mu_{{j'}D})\right]\\
	        &= \mathbb{E}_{\bD_{V\setminus\{j\}}\vert \bY_{V\setminus\{h\}}}\left[\Phi(-\kappa_{jj}^{1/2}\mu_{jD})\right]\\
	        &= \mathbb{P}(Y_j<0\vert \bY_{V\setminus\{j\}}),
	    \end{align*}
	    where $\Phi(.)$ is the cdf of standard univariate normal distribution. The fourth and the fifth equivalence hold since $\mu_{jD}=\mu_j - \kappa_{jj}^{-1}\sum_{v\in V\setminus\{j,j'\}} \kappa_{jv}(Y_v/d_v - \mu_v)$ does not depend on $(d_j,d_{j'},Y_j,Y_{j'})$.

        By comparing the conditional distribution of \eqref{eq:uniNorm} with the Equation (3), we can view the conditional distribution of \eqref{eq:uniNorm} as a regression model with dependent variable $Y_j/d_j$, the independent variable $Y_v/d_v$ and $\beta_{jv}=-\kappa_{jv}/\kappa_{jj}$ for every $v\in V\setminus\{j\}$. Obviously, $\beta_{jv}=0$ when $\kappa_{jv}=0$ implying the following conditional independence:
	    \begin{align*}
	        p(Y_j/d_j \vert \bY_{V\setminus\{j\}},Y_{j'},\bD) = p(Y_j/d_j \vert \bY_{V\setminus\{j\}},\bD).
	    \end{align*}
	    and the conditional sign independence:
	    \begin{align*}
	        \mathbb{P}(Y_j<0 \vert \bY_{V\setminus\{j\}},Y_{j'}) &= \mathbb{E}_{D\vert \bY_{V\setminus\{j\}},Y_{j'}}[\mathbb{P}(Y_j<0 \vert \bY_{V\setminus\{j\}},Y_{j'},\bD)]\\
	        &=\mathbb{E}_{D\vert \bY_{V\setminus\{j\}},Y_{j'}}[\Phi(-\kappa_{jj}^{1/2}\mu_{jD})]\\
	        &=\mathbb{E}_{D\vert \bY_{V\setminus\{j\}}}[\Phi(-\kappa_{jj}^{1/2}\mu_{jD})]\\
	        &=\mathbb{P}(Y_j<0 \vert \bY_{V\setminus\{j\}})
	    \end{align*}
	    Last, we include covariates in the model with functional coefficients $\beta(\bX)$. Assume the joint distribution of $\bY\bD$ follows a multivariate normal distribution with a mean zero and a functional precision depending on the covariates $\bX=[X_{1},\ldots,X_{q}]^\transp$. Specifically, the joint distribution can be written as 
	    \begin{align*}
            \bY\bD \mid \bD, \bX=\begin{bmatrix}Y_j/d_j \\ Y_{j'}/d_{j'} \\ \bY_{V\setminus\{j,{j'}\}}\bD_{V\setminus\{j,{j'}\}}
            \end{bmatrix} \sim 
            \bN_p\left(\mathbf{0}, 
            \begin{bmatrix}
                \kappa_{jj}(\bX) & \kappa_{j{j'}}(\bX) & \bkappa_{j.}(\bX)^\transp \\ 
                \kappa_{j{j'}}(\bX) & \kappa_{j'j'}(\bX) & \bkappa_{j'.}(\bX)^\transp \\
                \bkappa_{j.}(\bX) & \bkappa_{{j'}.}(\bX) & \cK_{V\setminus\{j,j'\}}(\bX)
            \end{bmatrix}^{-1}
            \right),
        \end{align*}
	    We can therefore have conditional distribution of $\frac{Y_j}{d_j}$ as:
	    \begin{align}\label{eq:CondPf_func}
	        \left. \frac{Y_j}{d_j} \right\vert \bY_{V\setminus\{j\}},\bD,\bX \sim N\left(\tilde{\mu}_{jD}(\bX), \kappa^{-1}_{jj}(\bX)\right),
	    \end{align}
	    where $\tilde{\mu}_{jD}(\bX)=-\frac{1}{\kappa_{jj}(\bX)}\sum_{v\in V\setminus\{j\} \cup \{{j'}\}} \kappa_{jv}(\bX)(Y_v/d_v)$. Comparing the conditional distribution of \eqref{eq:CondPf_func} and (3), 
        we can define the functional coefficients $\beta_{jv}(\bX)=-\kappa_{jv}(\bX)/\kappa_{jj}(\bX)$. Therefore, the covariance of the joint distribution becomes 
	    $\bSigma(\bX)=\begin{bmatrix}
	    \kappa_{11}(\bX) & \kappa_{12}(\bX) & \ldots & \kappa_{1p}(\bX)\\
	    \kappa_{12}(\bX) & \kappa_{22(\bX)} & \ldots & \kappa_{2p}(\bX)\\
	    \kappa_{13}(\bX) & \kappa_{32}(\bX) & \ldots & \kappa_{3p}(\bX)\\
	    \vdots & \vdots & \ddots & \vdots\\
	    \kappa_{1p}(\bX) & \kappa_{2p}(\bX) & \ldots & \kappa_{pp}(\bX)
	    \end{bmatrix}$. Following the derivation above, we replace the scalar $\beta$ by the functional coefficients $\beta(\bX)$ and have the bivariate normal as \eqref{eq:biNorm} with functional mean: 
	    \begin{align*}
	        \begin{bmatrix} \mu_{jD}(\bX) \\ \mu_{{j'}D}(\bX)\end{bmatrix} = 
	        -\begin{bmatrix}
	        \kappa_{jj}(\bX) & \kappa_{jj'}(\bX)\\
	        \kappa_{j{j'}}(\bX) & \kappa_{j'j'}(\bX)
	        \end{bmatrix}^{-1}
	        \begin{bmatrix}
	        \sum_{v\in V\setminus\{j,{j'}\}} \kappa_{jv}(\bX)Y_v/d_v \\ 
	        \sum_{v\in V\setminus\{j,{j'}\}} \kappa_{{j'}v}(\bX)Y_v/d_v 
	        \end{bmatrix}.
	    \end{align*}
	    When $\kappa_{jj'}(\bX)=0$, $\beta_{jv}(\bX)=-\kappa_{jv}(\bX)/\kappa_{jj}(\bX)=0$ implying the conditional independence with 
	    \begin{align*}
	        p(Y_j/d_j \vert \bY_{\textrm{pa}(j|\bX)},Y_{j'},\bD,\bX) = p(Y_j/d_j \vert \bY_{V\setminus\{j\}},\bD,\bX).
	    \end{align*}
	    and the corresponding conditional sign independence
	    \begin{align*}
	        \mathbb{P}(Y_j<0 \vert \bY_{V\setminus\{j\}},Y_{j'},\bX) 
	        &= \mathbb{E}_{D\vert \bY_{V\setminus\{j\}},Y_{j'},\bX} [\mathbb{P}(Y_j<0 \vert \bY_{V\setminus\{j\}},Y_{j'},\bD,\bX)]\\
	        &=\mathbb{E}_{D\vert \bY_{V\setminus\{j\}},Y_{j'},\bX}[\Phi(-\kappa_{jj}^{1/2}\mu_{jD}(\bX))]\\
	        &=\mathbb{E}_{D\vert \bY_{V\setminus\{j\}},\bX}[\Phi(-\kappa_{jj}^{1/2}\mu_{jD}(\bX))]\\
	        &=\mathbb{P}(Y_j<0 \vert \bY_{V\setminus\{j\}},\bX)
	    \end{align*}
\end{proof}



\section{Posterior Inference}\label{supp:Alg}
In this Section, we present the posterior inference procedure for rBGR including the MCMC algorithm and the symmetrization. We first provide details of the parameter expansion for the covariate coefficients $\alpha_{j,k,h}$ in Section \ref{supp:ParamExp}. Section \ref{supp:GibbsSmpr} describes the MCMC algorithm including the derivation of Gibbs sampler for the thresholded parameters. In Section \ref{supp:Symm}, we offer the rules used for symmetrizing both the covariate coefficients and the edges.
 
\subsection{Parameter Expansion}\label{supp:ParamExp}
In rBGR, we assign a spike-and-slab for covariate coefficients $\alpha_{j,k,h}$ with the parameter expansion technique \citep{BGR_Normal, doi:10.1080/01621459.2012.737742} to improve the mixing of MCMC. Let $\alpha_{j,k,h}=\eta_{j,k,h}\xi_{j,k,h}$. We impose a spike-and-slab prior on $\eta_{j,k,h}\sim N(0,s_{j,k,h})$ with  $s_{j,k,h}=\gamma_{j,k,h}\nu_{j,k,h}$, $\nu_{j,k,h}\sim InvGa(a_\nu, b_\nu)$, and $\gamma_{j,k,h}\sim \rho_j \delta_1(\gamma_{j,k,h}) + (1-\rho_j) \delta_{v_0}(\gamma_{j,k,h})$, where $v_0$ is a small pre-specified hyperparameter. Obviously, the prior results in a binary scenario in terms of $\gamma_{j,k,h}$. When $\gamma_{j,k,h}=v_0$ (spike),  $s_{j,k,h}$ is close to zero and results in negligible $\eta_{j,k,h}$ and $\alpha_{j,k,h}$ implying no effect from covariate $X_h$ on the edge between nodes $j$ and $k$. When $\gamma_{j,k,h}=1$ (slab), $\alpha_{j,k,h}$ is non-zero with a linear effect from $X_h$ on the edge between nodes $j$ and $k$. We then assign a beta distribution on $\rho_j \sim Beta(a_\rho,b_\rho)$. For $\xi_{j,k,h}$, we assign a mixture of two normal distribution, $\xi_{j,k,h} \sim N(m_{j,k,h},1)$ with $m_{j,k,h} \sim 0.5 \delta_1(m_{j,k,h}) + 0.5 \delta_{-1}(m_{j,k,h})$. The bimodal mixture distribution encourage $\alpha_{j,k,h}$ to be away from zero, which has been shown to improve selection \citep{doi:10.1080/01621459.2012.737742}.


\subsection{MCMC algorithm}\label{supp:GibbsSmpr}
At each iteration, the MCMC algorithm for rBGR updates parameters that consists of three parts: (i) thresholded parameters of $\alpha_{j,k,h}$ and $t_j$, (ii) random scales of $d_{ij}$, and (iii) hyperparameters. The closed-form of the full conditional distribution for thresholded parameters and hyper-parameters are available and enables the Gibbs sampler. On the other hand, we implement the Metropolis–Hastings algorithm for random scales. However, the derivation of the closed-form of the full conditional distribution for thresholded parameters is not straightforward. We briefly describe the general form of the thresholded parameters with Algorithm \ref{algo1} and refer to \citet{ThrhdMixDist} for more details. We then apply Algorithm \ref{algo1} to the thresholded parameters in rBGR. We summarize the whole MCMC algorithm in Algorithm \ref{algo_all}.


\paragraph{General algorithm for the thresholded parameter}
Consider a random variable $\theta$. Let $f_j(\theta)=a_{1j}\theta^2 + a_{2j}\theta + a_{3j}$ and $g_k(\theta) = b_{1k}\theta^2 + b_{2k}\theta + b_{3k}$. Consider the density of $\theta$ to be proportional to 
\begin{align*}
    \exp\left\{\sum_{j=1}^J f_{j}(\theta)\mathbbm{I}(\theta>L_j)+\sum_{k=1}^K g_{k}(\theta)\mathbbm{I}(\theta<U_k)\right\},
\end{align*}
where $L_j, j=1,\ldots,J$ are lower bounds for $f_j$ and $U_k, k=1,\ldots,K$ are upper bounds for $g_k$. We can classified $\theta$ into three different mixture distributions based on the values of coefficients in $f_j$ and $g_k$:
\begin{itemize}\setlength\itemsep{-0.5em}
    \item[1.] If at least one of $\{a_{1j},\ldots,a_{1J},b_{1k},\ldots,b_{1K}\}$ is non-zero, then $\theta$ follows a mixture of truncated normal distributions.
    \item[2.] If $a_{1j}=b_{1k}=0,\forall j,k$ and at least one of $\{a_{2j},\ldots,a_{2J},b_{2k},\ldots,b_{2K}\}$ is non-zero, then $\theta$ follows a mixture of exponential distributions.
    \item[3.] If $a_{1j}=b_{1k}=a_{2j}=b_{2k}=0,\forall j,k$ and at least one of $\{a_{3j},\ldots,a_{3J},b_{3k},\ldots,b_{3K}\}$ is non-zero, then $\theta$ follows a mixture of uniform distributions.
\end{itemize}
The key idea is to exhaust the real line into mutually exclusive intervals and update the random variable $\theta$ within each interval. We start by dissecting the real line into $J+K+1$ intervals using the lower or upper bounds as endpoints. For each interval, the truncation mechanism for all functions of $f_j$ and $g_k$ is determined, and we only need to consider the coefficients from non-zero functions of $f_j$ and $g_k$. With the given coefficients, we can easily derive the distribution within each interval. Finally, we collect distributions from all intervals and normalize the distribution. We implement this idea in Algorithm \ref{algo1}.

\begin{algorithm}[!htb]
    \caption{Full Condition for $\theta$}\label{algo1}
    \hspace*{\algorithmicindent} \textbf{Input}: 
    \vspace*{-3mm}
    \begin{itemize} \itemsep -0.2em
        \item[(a)] $\{L_j\}_{j=1}^J, \{U_k\}_{k=1}^K, \{f_j(\theta)=a_{1j}\theta^2 +a_{2j}\theta+a_{3j} \}_{j=1}^J$ and $\{g_k(\theta)=g_k(\theta) = b_{1k}\theta^2 + b_{2k}\theta + b_{3k}\}_{k=1}^K$.
        \item[(b)] The prior on $\theta$ with the kernel $\exp\left(c_1\theta^2 + c_2\theta +c_3\right)$.
    \end{itemize}
    \hspace*{\algorithmicindent} \textbf{Output}: 
    The full condition distribution of $\theta$.
    \begin{algorithmic}[1]
    \State Sort the bounds of $\{L_1,\ldots,L_J, U_1,\ldots, U_K\}$ in ascending order with $J+K+1$ intervals dissected from the real line $\mathbb{R}=\cup_{i=1}^{J+K+1} \cI_i$.
        \For{Each interval $\cI_i, i = 1...J+K+1$}
            \State Initialize $D_i=c_1, E_i=c_2$ and $F_i=c_3$.
            \For{$j = 1...J, k=1,\ldots,K$} 
                \If {$\cI \subset [L_j,\infty)$}
                \State Update $D_i=D_i + a_{1j}, E_i=E_i + a_{2j}$ and $F_i=F_i + a_{3j}$.
                \EndIf
                \If {$\cI \subset (-\infty,U_k]$}
                \State Update $D_i=D_i + b_{1j}, E_i=E_i + b_{2j}$ and $F_i=F_i + b_{3j}$.
                \EndIf
            \EndFor
            \If{$D_i\neq 0$}
                \State $\theta \sim N_{\cI_i}(-\frac{E_i}{2D_i},-\frac{1}{D_i})$ for $\theta\in \cI_i$.
            \EndIf
            \If{$D_i= 0$ and $E_i \neq 0$}
                \State $\theta \sim Exp_{\cI_i}(E_i)$ for $\theta\in \cI_i$.
            \EndIf
            \If{$D_i=E_i=0$ and $F_i\neq 0$}
                \State $\theta$ follows a uniform distribution on $\cI_i$.
            \EndIf
        \EndFor 
        \State Normalize the whole distribution $\theta$, which is proportional to $\sum_{i=1}^{J+K+1} M_ih_i(\theta)$ and $M_i$ is the normalizing constant independent of $\theta$ for the distribution $h_i(\theta)$ on interval $\cI_i$.
    \end{algorithmic}
\end{algorithm}

From Algorithm \ref{algo1}, it is obvious that the conjugacy of $\theta$ can be achieved by assigning priors for different values for $f_j(\theta)$ and $g_k(\theta)$. Specifically, when $a_{1j}=a_{2j}=b_{1k}=b_{2k}=0$ for all $j=1,\ldots,J$ and $k=1,\ldots,K$, we can assign a uniform prior with $c_1=c_2=0$ and $c_3\neq0$ resulting in a mixture of uniform distribution. Since the prior is a special case of the mixture of uniform distribution with only one component, the conjugacy is attainable. Meanwhile, if we assign a normal prior with $c_1\neq0$, we obtain a mixture of truncated normal, which grants the conjugacy for $\theta$ with normal prior. Given the Algorithm, we then derive the full condition distribution for thresholded parameters from rBGR with the Gibbs sampler.


\paragraph{Covariate coefficients.}
We first derive the full condition for $\eta_{j,k,h}$ and $\xi_{j,k,h}$. We only show the full condition for $\eta_{j,k,h}$ since both are normally distributed, and the distribution of $\xi_{j,k,h}$ can be analogously derived.

\begin{align}\label{eq:GibbsCovCoef}
    \nonumber p(\eta_{j,k,h}\mid \bY,\bX,\Theta_{-\eta_{j,k,h}}) &\propto
    \exp\left\{\sum_{i:X_{ih}\geq 0} \left[g_i(\eta_{j,k,h}) \mathbbm{I}(\eta_{j,k,h}\geq T_{i1}) + g_i(\eta_{j,k,h}) \mathbbm{I}(\eta_{j,k,h}<T_{i2}) \right] \right. \\
    &+ \left. \sum_{i:X_{ih}<0}\left[g_i(\eta_{j,k,h}) \mathbbm{I}(\eta_{j,k,h}\geq T_{i2}) + g_i(\eta_{j,k,h}) \mathbbm{I}(\eta_{j,k,h}<T_{i1}) \right] \right\}
   \nonumber g_i(\eta_{j,k,h}) \\
   &= a_{1i}\eta_{j,k,h}^2 + a_{2i}\eta_{j,k,h} \\
   \nonumber a_{1i} &= -\frac{X_{ih}^2 Y_{ik}^2\xi_{j,k,h}^2}{2\sigma^2_jd_{ik}^2} - \frac{1}{2s_{j,k,h}}\\ 
   \nonumber a_{2i} &= -\frac{X_{ih}\xi_{j,k,h}}{\sigma^2_j}\left[\frac{Y_{ik}^2}{d_{ik}^2}\sum_{l\neq h}^q\alpha_{j,k,l}X_{il} + \frac{Y_{ik}}{d_{ik}}\left( -\frac{X_{ij}}{d_{ij}} + \sum_{m\neq k}^p \beta_m(\bX_i)Y_{im} \right) \right]\\
   \nonumber T_{i1} &= \frac{t_j - \sum_{l\neq h}^q \alpha_{j,k,l}X_{il}}{\xi_{j,k,h}X_{ih}};\thickspace T_{i2}=\frac{-t_j-\sum_{l\neq h}^q \alpha_{jl}X_{il}}{\xi_{j,k,h}X_{ih}}
\end{align}
where $g_i(\eta_{j,k,h})= a_{1i}\eta_{j,k,h}^2 + a_{2i}\eta_{j,k,h}$ is a quadratic function of $\eta_{j,k,h}$ and $T_{i1}$ and $T_{i2}$ are independent of $\eta_{j,k,h}$. Therefore, the full condition distribution of $\eta_{j,k,h}$ belongs to the first category with a mixture of normal distribution. When we assign a normal prior on $\eta_{j,k,h}$ and $\xi_{j,k,h}$, we obtain the conjugacy with the Gibbs sampler shown in Algorithm \ref{algo1}.

\paragraph{Threshold parameter.}
The same idea can be used on the threshold parameter $t_j$. Specifically, the full condition of the threshold parameter is 
\begin{align}\label{eq:Gibbst}
    p(t_j\mid \bY,\bX,\Theta_{-t_j})
    &\propto \exp \left\{
    -\sum_{i=1}^n \sum_{k\neq j}^p \mathbb{I}\left(t_j<\lvert \theta_{j,k}(\bX_i) \rvert \right) \frac{P_{ik} + Q_{ik}}{2\sigma^2_j}\right\} \frac{1}{t_{\max}} \mathbb{I}(0\leq t_j\leq t_{\max}),
\end{align}
where $Q_{ik}= 2\theta_{j,k}(\bX_i)\frac{Y_{ik}}{d_{ik}}\sum_{k'\neq k}\theta_{j,{k'}}(\bX_i)\frac{Y_{i{k'}}}{d_{i{k'}}}\mathbb{I}\left(t_j<\lvert \theta_{j,{k'}}(\bX_i) \rvert \right)$ and $P_{ik}=\theta_{j,k}^2(\bX_i) \frac{Y_{ik}^2}{d_{ik}^2} - 2\theta_{j,k}(\bX_i)\frac{Y_{ik}}{d_{ik}}\frac{Y_{ij}}{d_{ij}}$. Given all $\theta_{j,k}(\bX_i)$, we claim that both $P_{ik}$ and $Q_{ik}$ are constant with respect to $t_j$. Obviously, $P_{ik}$ does not depend on $t_j$. For any given interval, we also find that $Q_{ik}$ is independent of $t_j$. Therefore, $Q_{ik}$ is also independent of $t_j$, and the full condition for $t_j$ falls into the third category with the mixture of the uniform distribution.

Now, we can present the whole MCMC algorithm as follows:
\begin{algorithm}[!htb]
    \caption{MCMC algorithm for rBGR}\label{algo_all}
    \vspace*{-3mm}
    \begin{itemize}
    \itemsep -0.2em
        \item[(a)] Update $\eta_{j,k,h}$ and $\xi_{j,k,h}$ by Algorithm \ref{algo1} with \eqref{eq:GibbsCovCoef};
        \begin{itemize}
            \item Rescale $\eta_{j,k,h}$ and $\xi_{j,k,h}$ with $\eta_{j,k,h}\rightarrow \eta_{j,k,h} \lvert \xi_{j,k,h} \rvert$ and $\xi_{j,k,h} \rightarrow \eta_{j,k,h}/\vert \xi_{j,k,h}\rvert$.
        \end{itemize}
        \item[(b)] Update $t_{j}$ by Algorithm \ref{algo1} with \eqref{eq:Gibbst};
        \item[(c)] Update $m_{j,k,h}$ by Gibbs: $p(m_{j,k,h}=1\mid \xi_{j,k,h})=\frac{1}{1+\exp(-2\xi_{j,k,h})}$;
        \item[(d)] Update $\gamma_{j,k,h}$ by Gibbs: $\frac{p(\gamma_{j,k,h}=1\mid \eta_{j,k,h},\nu_{j,k,h},\rho_{j,k,h})}{p(\gamma_{j,k,h}=v_0\mid \eta_{j,k,h},\nu_{j,k,h},\rho_{j})}=\frac{\sqrt{v_0}\rho_j}{1-\rho_j}\exp(\frac{-v_0 \eta_{j,k,h}^2}{2v_0\nu_{j,k,h}})$;
        \item[(e)] Update $\nu_{j,k,h}$ by Gibbs: $p(\nu_{j,k,h}\mid \eta_{j,k,h},\gamma_{j,k,h})\textrm{InvGa}(a_\nu +1/2, b_\nu + \frac{\eta_{j,k,h}^2}{2\gamma_{j,k,h}})$;
        \item[(f)] Update $\rho_{j}$ by Gibbs: $p(\rho_{j}\mid \gamma_{j,k,h})=\textrm{Beta}(a_\rho + \sum_{k,h}\mathbb{I}(\gamma_{j,k,h}=1),b_\rho+\sum_{k,h}\mathbb{I}(\gamma_{j,k,h}=v_0))$;
        \item[(g)] Update $d_{ij}$ by MH algorithm with a proposal as prior.
        \item[(h)] Update $\pi_{j}$ by Gibbs: $p(\pi_j\mid D_{ij})=\textrm{Beta}(a_\pi+\sum_{i}\mathbb{I}(D_{ij=1}), b_\pi+\sum_{i}\mathbb{I}(D_{ij\neq 1}))$;
    \end{itemize}
\end{algorithm}

\subsection{Details of Covariate and Edge Selection}\label{supp:Symm}
The estimated coefficients from rBGR of (3) do not guarantee the symmetry required in undirected graph. Moreover, due to the introduction of random factors with the CSI characterization, we only focus on the sign of the edge. In this section, we describe algorithms to symmetrize the estimated covariate coefficients $\hat{\alpha}_{j,k,h}$ and the sign of graph edges of $\hat{\beta}_{j,k}(\bX_i)$. Denote $\PP^\alpha_{j,k,h}=\PP(\hat{\alpha}_{j,k,h}\neq 0)$ as the posterior inclusion probability (PIP) of $\hat{\alpha}_{j,k,h}$ and let $\tilde{\alpha}_{j,k,h}$ be the covariate coefficients for the undirected graph between node $j$ and $k$ for covariate $h$. We formulate the symmetrization rules via choosing the direction with a lower PIP:
\begin{align}
    \label{eq:alphaMin}\tilde{\alpha}_{j,k,h} = \hat{\alpha}_{j,k,h}\mathbb{I}(\PP^\alpha_{k,j,h}>\PP^\alpha_{j,k,h}) +  \hat{\alpha}_{k,j,h}\mathbb{I}(\PP^\alpha_{j,k,h}\geq\PP^\alpha_{k,j,h}).
\end{align}
Given a cutoff $c_0$, Equation \ref{eq:alphaMin} requires both directions to have PIPs bigger than $c_0$ implying a network with less edges. Another possible symmetrization is
\begin{align}
    \label{eq:alphaMax}\tilde{\alpha}_{j,k,h} = \hat{\alpha}_{j,k,h}\mathbb{I}(\PP^\alpha_{j,k,h}>\PP^\alpha_{k,j,h}) +  \hat{\alpha}_{k,j,h}\mathbb{I}(\PP^\alpha_{k,j,h}\geq\PP^\alpha_{j,k,h}).
\end{align}
Obviously, Equation \eqref{eq:alphaMax} is less conservative and requires at least one PIP bigger than $c_0$. Similar symmetrization rules can be seen in \citet{RegGMM} if we replace the PIP with the absolute value of coefficients. For rules with absolute value of coefficients, both rules are asymptotically equivalent \citep{DAG_NN, RegGMM}, but the rule of \eqref{eq:alphaMin} performs better given finite samples \citep{DAG_NN}. We use the rule \eqref{eq:alphaMin} for the rest paper.

For the edge $\hat{\beta}_{j,k}(\bX_i)$, we first calculate the estimated linear function of $\tilde{\theta}_{j,k}(\bX_i)$ and symmetrize the edge posterior probability (ePP) of the sign of $\hat{\beta}_{j,k}(\bX_i)$. Specifically, $\tilde{\theta}_{j,k}(\bX_i)=\sum_{h=1}^q \tilde{\alpha}_{j,k,h} X_{ih}$ and $\tilde{\theta}_{j,k}(\bX_i)$ is symmetric since $\tilde{\alpha}_{j,k,h}$ from \eqref{eq:alphaMin} is symmetric. Denote $\PP^\beta_{j,k}(\bX_i)=\PP(\hat{\beta}_{j,k}(\bX_i)\neq0)$ as the ePP of a directed edge from node $k$ to $j$. In this paper, we symmetrize the sign of the edge by taking the maximum of the ePP from two directions through 
\begin{align}
    \label{eq:betaMax}\tilde{\PP}^\beta_{j,k}(\bX_i) = \max(\PP^\beta_{j,k}(\bX_i), \PP^\beta_{k,j}(\bX_i)),
\end{align}
where $\tilde{\PP}^\beta_{j,k}(\bX_i)$ is ePP of an undirected edge between node $j$ and $k$. Given a threshold $c_1$, we then call an undirected edge if $\tilde{\PP}^\beta_{j,k}(\bX_i)>c_1$. Alternatively, we can take the minimum as 
\begin{align}
    \label{eq:betaMin}\tilde{\PP}^\beta_{j,k}(\bX_i) = \min(\PP^\beta_{j,k}(\bX_i), \PP^\beta_{k,j}(\bX_i)).
\end{align}
Clearly, \eqref{eq:betaMin} is more conservative and needs both $\PP^\beta_{j,k}(\bX_i)$ and $\PP^\beta_{j,k}(\bX_i)$ bigger than $c_1$ to call an edge, while \eqref{eq:betaMax} requires only one of the posterior probability bigger than $c_1$. 

Once we symmetrize the ePP, we can decide the sign for edges given that the ePP is bigger than the cutoff $\tilde{\PP}^\beta_{j,k}(\bX_i)>c_1$. Without loss of generality, assume that we chose a specific direction as undirected edge with $\tilde{\PP}^\beta_{j,k}(\bX_i)=\PP^\beta_{j,k}(\bX_i)$. We estimate the sign of the edge by comparing the posterior probability of positive and negative for the chosen direction. Specifically, given the direction of $\tilde{\PP}^\beta_{j,k}(\bX_i)=\PP^\beta_{j,k}(\bX_i)$, we estimate the sign of the edge by the following rule:
\begin{align}\label{eq:edgeSign}
    \textrm{sign}(\beta_{j,k}(\bX_i))=\begin{cases}1 & \textrm{ if } \PP(\hat{\beta}_{j,k}(\bX_i)> 0) > \PP(\hat{\beta}_{j,k}(\bX_i)< 0) \\  -1 
    & \textrm{ if } \PP(\hat{\beta}_{j,k}(\bX_i)>0) \leq \PP(\hat{\beta}_{j,k}(\bX_i)< 0) \end{cases}
\end{align}

\begin{remark}
Both rules of \eqref{eq:betaMax} and \eqref{eq:betaMin} leave the value of $\hat{\beta}_{j,k}(\bX_i)$ to be asymmetric. One might symmetrize edges through symmetrizing both linear function and the threshold parameter. However, matching the threshold parameter results in a common $\hat{t}_j=\hat{t}$ for all $j=1,\ldots,p$, which imposes strict constraints. For this paper, we do not require the value of $\hat{\beta}_{j,k}(\bX_i)$ from two directions equal and only need to ensure that the sign of edges from two directions agrees. 
\end{remark}

\section{Additional Results for Simulation Studies}\label{supp:AddSimulation}
\subsection{Details of Data Generating Mechanism}\label{supp:SimDataGen}
We generate the data from an underlying multivariate normal distribution with precision matrix representing the undirected graph and transform the latent normal data with random scale to obtain the observed non-normal data. Specifically, we first generate the covariates $\bX_i \overset{iid}{\sim} U(-1,1)$ and obtain the latent data from a multivariate normal distribution. By multiplying the latent data by random scales, we acquire the observed non-normal data. We set the sample size and the dimension of $\bY_i$ and $\bX_i$ as $(n,p,q)=(250,50,3)$, and generate the latent data from the following procedures:
\begin{align*}
	 \bY_i^* = \left[Y_{i1}^*,\ldots,Y_{ip}^*\right]^\transp \overset{iid}{\sim}\bN_p(\mathbf{0},\bOmega^{-1}(\bX_i)), i=1,\ldots,n
\end{align*}
where $\bOmega^{-1}(\bX_i)$ is the true precision matrix. For true precision matrix, we assign unit diagonal elements and randomly pick $2\%$ of the off-diagonal to be non-zero. Given a threshold parameter $t^0$, each non-zero precision depends linearly on the covariates and is truncated to zero if the absolute value is smaller than the threshold parameter $t^0$. Specifically, we set the non-zero precision as $\omega^{j,k}(\bX_i) = r^{j,k}(\bX_i) \mathbb{I}(\lvert r^{j,k}(\bX_i)\rvert >t^0)$ and $r^{j,k}(\bX_i) =\sum_{h=1}^q X_{ih} \nu_{j,k,h}$, where $\nu_{j,k,h}\sim U(-0.5,-0.35)\cup U(0.35,0.5)$. We set $t^0=0.15$ to filter around half of the non-zero off-diagonal elements. The final precision matrix might not be positive semi-definite, and we repeat the whole process till the precision matrix is positive semi-definite. We obtain the random scales from a mixture distribution of the point mass at one and a inverse gamma distribution with shape and scale parameters $d^2_{ij} \overset{iid}{\sim} (1-\pi)\delta_1 +\pi InvGa(a_{d_{j}},b_{d_{j}})$. We assign three different levels of $\pi$ representing three different levels of non-normal contamination: $\pi\in\{0,0.5,0.8\}$. Given the latent data from the multivariate normal distribution, we multiply the random scales, $d_{ij}$, to generate the observed data of $[Y_{i1},\ldots, Y_{ip}]=[Y_{i1}^*d_{i1},\ldots,Y_{ip}^*d_{ip}]$.

\subsection{Convergence of MCMC}
One important issue for the Bayesian method is to ensure that the MCMC converged to draw the samples from the target posterior distribution. We investigate the convergence of the MCMC through the Geweke statistics \citep{Geweke92evaluatingthe}. Specifically, we check the Geweke statistics of the covariate coefficients $\alpha_{j,k,h}$. After the burn-in period, we take the first and the last $20\%$ of the posterior samples and calculate the Geweke statistics. We require p-values for all $\alpha_{j,k,h}$ to be insignificant after the Bonferroni correction \citep{pmid24697967} to ensure the convergence of the algorithm.

\subsection{Simulation Results of Different cut-off of $c_0$ and $c_1$ Controlling for False Discovery Rates}
Another possible way to decide the cut-off of $c_0$ and $c_1$ is by controlling the false discovery rate (FDR) \citep{FDR_Paper} $\alpha$. Consider a sorted vector $Q$ of dimension $N$ in decreasing order with each element as a probability. Denote $Q_{(k)}$ as the $k$-th largest element in $Q$. We first calculate $\xi=\max\{K:K^{-1}\sum_{k=1}^K(1-Q_{(k)}) <\alpha\}$ and set the cut-off as $c^\alpha=Q_{(\xi)}$.  In this Section, we fixed the FDR at $\alpha=0.1$ and obtain the cut-off for the PIP from $\alpha_{j,k,h}$ and the ePP from $\beta_{jk}(\bX_i)$. 

\begin{figure}
    \centering
    \includegraphics[width=\textwidth]{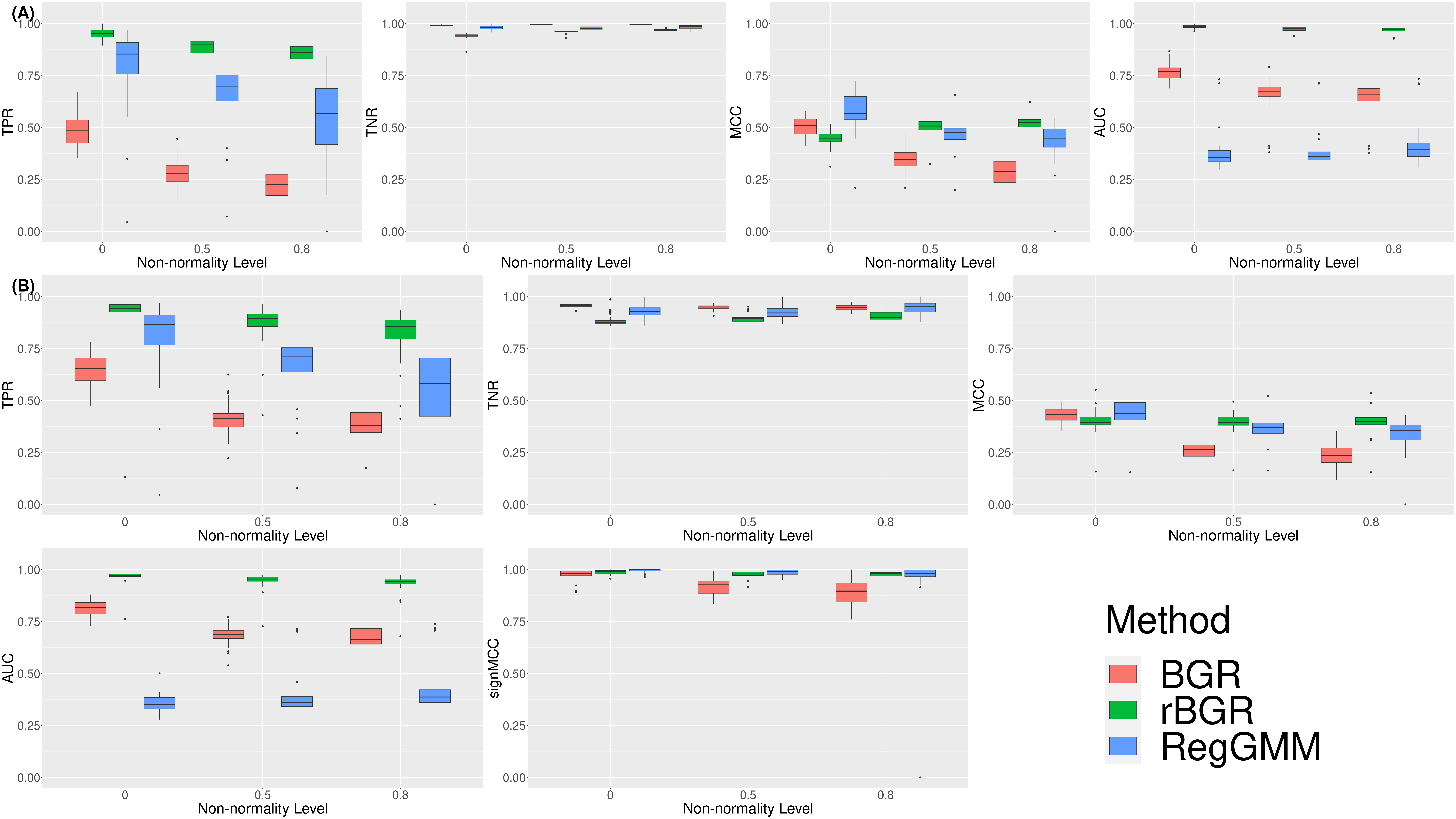}
    \caption{Graph recovery for BGR (red), rBGR (green) and RegGMM (blue) under different levels of non-normality in terms of (A) covariates selection (top row) and (B) edge selection (bottom two rows). Panel (A) measures the covariate selection through four metrics (from left to right: TPR, TNR, MCC and AUC) are measured under three different levels of non-normality. Panel (B) demonstrates the edge selection by four criteria (from upper left to lower right: TPR, TNR, MCC, AUC) and the sign consistency by sign-MCC (lower left) for non-zero edges. All values for TPR, TNR and MCC are measured at a cut-off controlling for false discovery rate.}
    \label{sfig:sim_FDR}
\end{figure}

Panel (A) of Figure \ref{sfig:sim_FDR} show the results for covariate selection when we use the cut-off controlling for the false discovery rate. Comparing to the cut-off at $c_0=0.5$ used in Main Paper, we observe that rBRG and BGR generate a higher TPR and TNR but a lower MCC for covariate selection. Specifically, rBGR outperforms both BGR and RegGMM in TPR across different non-normality levels. For TNR, rBGR performs slightly worse than BGR and RegGMM for across all non-normality levels, but the disadvantage of rBGR decreases when the non-normality level increases. Moreover, all three methods select correct covariates and edges ($>90\%$) with small difference ($<10\%$) in terms of TNR. We observe that rBGR achieves a lower MCC comparing to BGR and RegGMM when the data is normally distributed. However, rBGR surpasses BGR and RegGMM in terms of MCC when the level of non-normality increases. Similar to Main Paper, modeling the non-normality from random scales in rBGR is favored compared to models without random scales in terms of covariate selection. 

We show the graph recovery for the edge selection using the cut-off controlling for the false discovery rate in Panel (B) of Figure \ref{sfig:sim_FDR}. We observe that using the cut-off controlling for the false discovery rate results in a higher TPR, but lower TNR and MCC for rBGR. Specifically, rBGR has the best performance in terms of TPR comparing to other benchmarks of BGR and RegGMM under all levels of non-normality, and the advantage of rBGR becomes more prominent as the non-normality increases. For MCC, rBGR is slightly inferior than BGR and RegGMM under the normal distribution. However, rBGR is favored when the non-normality level increases. Both TNR and sign-MCC show excellent edge selection performance ($>90\%$) for all three methods, with minimal differences ($<10\%$) across the three non-normality levels. In summary, modeling the non-normality through random scales in rBGR result in equivalent (under normal distribution) or better performances in all metric for edge selection compared to the other methods without accounting for non-normality.

\section{Additional Results for Real Data Analysis}\label{supp:RealDataRes}

\subsection{Pre-processing Procedures and Convergence}\label{supp:PreProcessConv}
\begin{figure}
    \centering
    \includegraphics[width=\textwidth]{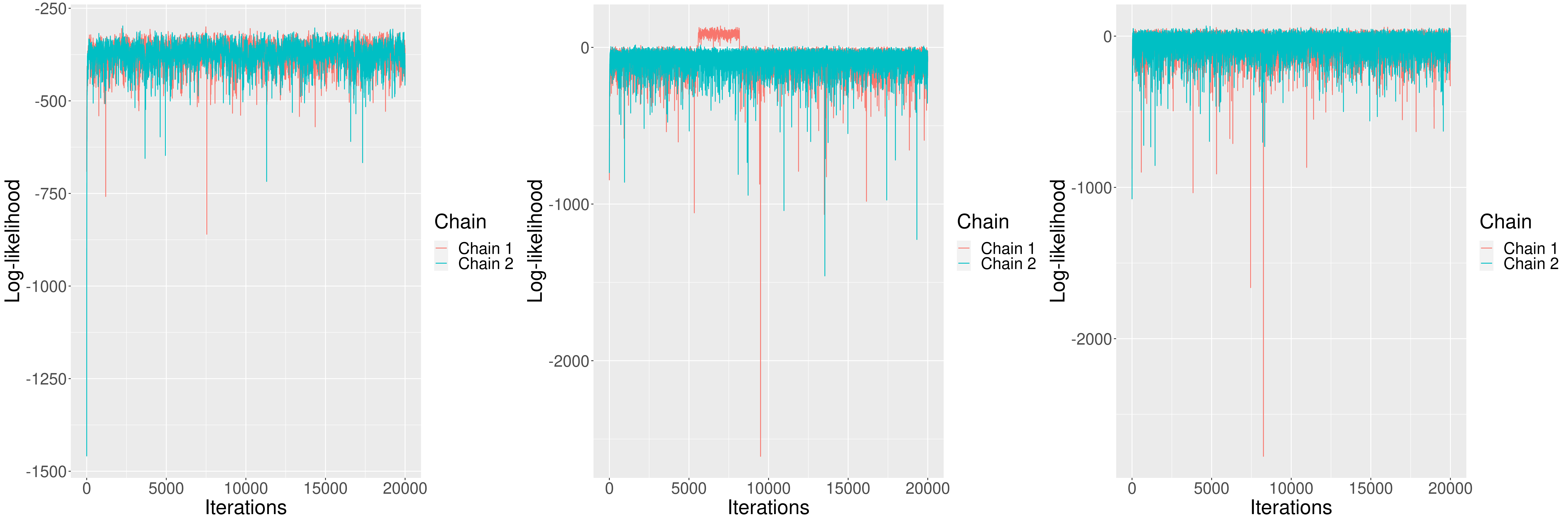}
    \caption{Convergence diagnostics for using rBGR algorithm on lung cancer. Three randomly chosen nodes are initiated with two different chains. Both chains converge to a similar level of log-likelihood after the burn-in period of the first $19,000$ iterations.}
    \label{sfig:rBGR_ChkConv_llh}
\end{figure}

For proteomics data, we first removed phosphorylation proteins and focus on proteins in 12 important cancer-related pathways (apoptosis, breast hormone signaling, breast reactive, cell cycle, core reactive, DNA damage response, EMT, PI3K/AKT, RAS/MAPK, RTK, TSC/mTOR and hormone receptor) \citep{PRECISE}. After centering proteomic data, we obtain 41 proteins from both OV and LUAD with 241 patients and 360 patients for OV and LUAD, respectively. For covariates, we obtained expression data from immune cells and treated the mRNA expression as the immune cell abundance. We averaged mRNA expression for the genes listed for seven immune cells (B cell, T cell, macrophages, monocytes, neutrophils, natural killer cells and plasma cell) and three pathways (proliferation, interferon and translation) \citep{ImSig}. We further took the log transformation and standardized on the averaged expression data. For this analysis, we chose T cells and two important components of myeloid-derived suppressor cells (MDSC), monocytes and neutrophils, for both OV and LUAD for two reasons. First, both T cells and MDSC are essential in both OV \citep{pmid34248988, pmid32850861} and LUAD \citep{pmid33494181, pmid36008393}. The existing biology also suggests the importance of macrophage and natural killer cells (NK cells), but since we  observed a high correlation among T cells, macrophages (OV: 0.71 and LUAD: 0.80) and NK cells (OV: 0.77 and LUAD: 0.49) we did not include the marcophages and NK cells in this analysis. We ran rBGR on OV and LUAD with $20,000$ iterations and discarded first $19,000$ iterations. We adapted the symmetrization of \eqref{eq:alphaMin} for covariate coefficients and \eqref{eq:betaMax} for edges. We examine the convergence of the algorithm through both the Geweke statistics and the likelihood trace plot shown in Figure \ref{sfig:rBGR_ChkConv_llh}. Specifically, we ensure the convergence of the algorithm by requiring the p-values of $\alpha_{j,k,h}$ from the Geweke statistics are all insignificant after Bonferroni correction \citep{pmid24697967}. In Figure \ref{sfig:rBGR_ChkConv_llh}, we randomly pick three proteins and run the algorithm with two chains of different initialization. Both chains converge to a similar level of log-likelihood after the burn-in period of the first $19,000$ iterations, indicating the convergence of the algorithm.

\subsection{Patient-Specific Networks for Ovarian Cancer}
\begin{figure}
    \centering
    \includegraphics[width=\textwidth]{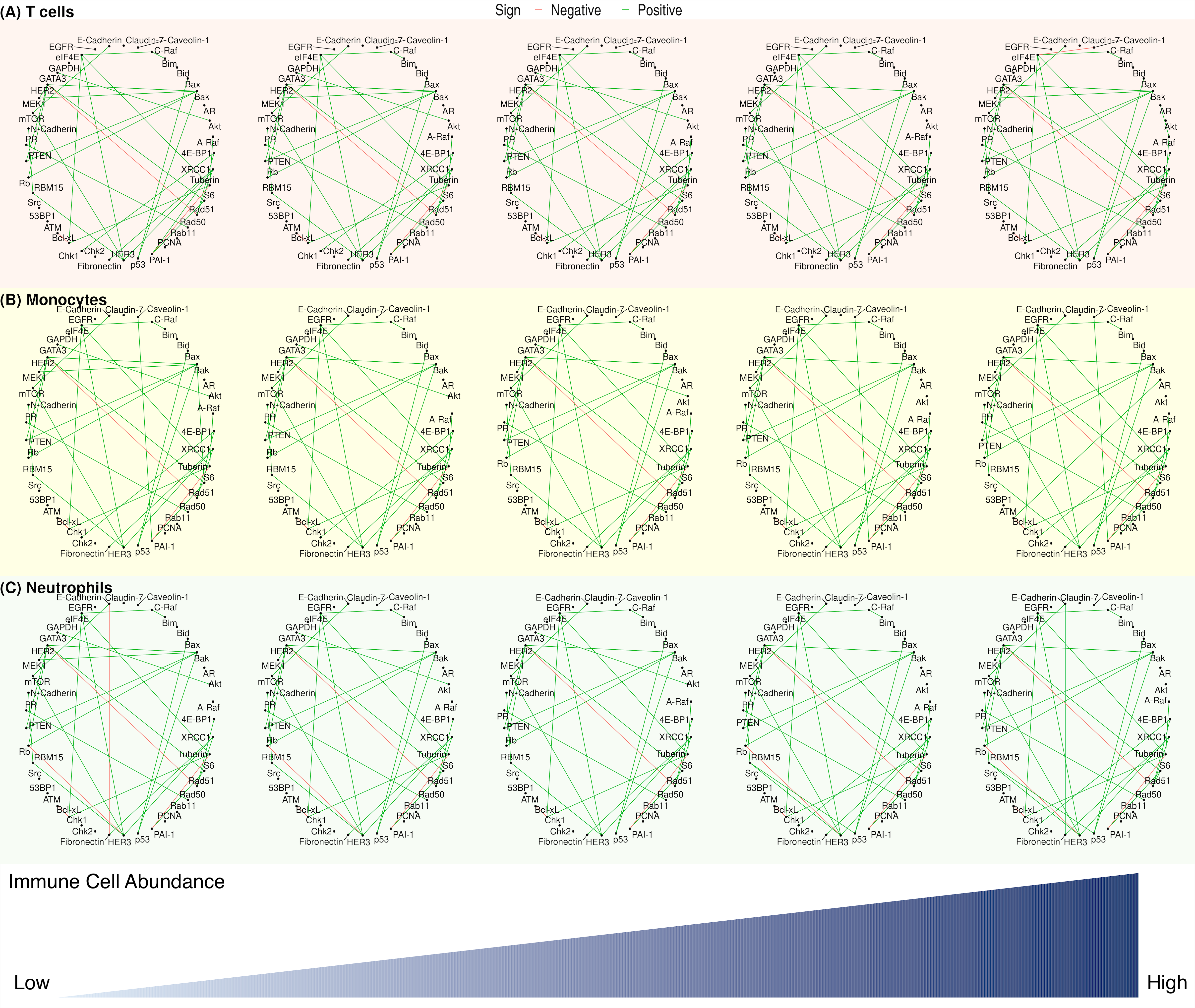}
    \caption{Networks of OV under five different percentiles immune component of (A) T cells, (B) monocytes and (C) neutrophils with the rest two components fixed at mean zero. The estimated network for varying immune components are shown from the left to right for 5, 25, 50 ,75, and 95-th percentiles. Edges are identified with signs (green: positive and red: negative) when the ePPs are bigger than $c_1=0.5$.}
    \label{sfig:indNet_OV}
\end{figure}

In this Section, we present the patient-specific network for ovarian cancer (see Figure \ref{sfig:indNet_OV}). Similar to the Main Paper, we vary the abundance of one immune component with the rest two components fixed and focus on the edges that change the sign when the immune component abundance increase. In OV, we observe that only the edge of E-Cadherin-Fibronectin changes the sign the neutrophils abundance increases. Specifically, the this edge is positively correlated to the neutrophil abundance. When neutrophil abundance is higher, E-Cadherin-Fibronectin is positive; vice-versa, E-Cadherin-Fibronectin is negative when neutrophil is scarce. Recently, neutrophils have been shown to induce the expression of fibronectin through the epithelial–mesenchymal transition pathway, and the same pathway also represses the expression of E-Cadherin, resulting in the tumor growth \citep{pmid32545405}. 

\section{rBGR Code Implementation}\label{supp:CodeImp}
We implemented rBGR with a general purpose \texttt{R} package that is available in \url{https://github.com/bayesrx/rBGR}. \texttt{rBGR} package allows users to construct graphs that vary based on subject-specific information when the normality assumption fails in the observed dataset. We briefly introduce the package with a manual and replicate our real data analysis in Main Paper as an example.

\subsection{Obtaining rBGR R package and Data}
The \texttt{rBGR} accommodates the non-normality by the random scales and builds graphs through graphical regressions. The coefficients of graphical regression incorporate the subject-specific information and encode the graph edges by the zero coefficients. Due to the formulation of graphical regression, \texttt{rBGR} obtains posterior samples of coefficients by Gibbs sampler and infers the random scales by the Metropolis-Hasting algorithm. We refer to more details in Main Paper Section 4 and Section \ref{supp:Alg}. 

In the package, we wrap the MCMC algorithm in a function of \texttt{rBGR\_mcmc\_Int()}. Given proteomic expression data and the subject-specific information (immune component abundance in the Main Paper), \texttt{rBGR} regresses one variable on the rest of the variables. Given the covariates, users need to execute the same function on all variables to construct the whole graph. Fortunately, this algorithm can be run parallelly to reduce the computation time. 
 
The MCMC function \texttt{rBGR\_mcmc\_Int()} takes several arguments with different options for users to control the algorithm. To run the MCMC, users are required to specify the data by the following three arguments: (i) regressand by the argument $Y$, (ii) regressor by the argument $X$, and (iii) the covariate by the argument $U$. The function \texttt{rBGR\_mcmc\_Int()} also allows additional options to control the model, such as $N$ for the number of iterations, \texttt{burnin} for the number of iterations to be discarded, and \texttt{seed\_} for the initial seeds. 

The function \texttt{rBGR\_mcmc\_Int()} produces the posterior samples of (i) two components of coefficients $\xi_{j,k,h}$ and $\eta_{j,k,h}$, (ii) threshold parameter $t_j$ and random scales $d_{ij}$. Given the covariates, users can obtain the posterior coefficients by $\alpha_{j,k,h}=\eta_{j,k,h}\xi_{j,k,h}$ and edges by $\beta_{j,k}(\bX_i)=\theta_{j,k}(\bX_i) \mathbf{I}(\lvert \theta_{j,k}(\bX_i) \rvert >t_j)$, where $\theta_{j,k}(\bX_i)=\sum_{h=1}^q \alpha_{j,k,h}X_{ih}$.
 
\subsection{Reproduciblility of Results}
Once we obtain posterior samples of coefficients and edges, users can symmetrize and summarize the results to obtain undirected graphs in two different levels: population ($\alpha_{j,k,h}$) and individual ($\beta_{j,k}(\bX_i)$) levels. We offer codes in the package to demonstrate our symmetrizing and summarizing algorithm in Section \ref{supp:Symm} with the data used in Main Paper Section 6. 

For the population-level graph, users can run \texttt{pstSmpExt\_pop.R} to extract the population-level information and visualize the results through the \texttt{PlotRes\_popLevel.R}. Similarly, in the individual-level graph, we first obtain the symmetrized edges by executing \texttt{pstSmpExt\_ind.R} and then visualize the results by \texttt{PlotRes\_indQuantile.R}. Currently, we use the 5, 25, 50, 75, and 95-th percentiles of the covariates as five different individuals, as shown in the Main Paper.

\clearpage
\FloatBarrier
\bibliography{rBGR}